\documentclass[amsmath, aps]{revtex4}
\usepackage[pdftex]{graphicx}
\usepackage{times}
\usepackage{epsfig}
\usepackage{amssymb}

\usepackage{amsmath}
\begin{document}
\title{QED theory of the multiphoton cascade transitions in atoms}
\author{L. Labzowsky$^{1,2}$, D. Solovyev$^1$ and T. Zalialiutdinov$^{1}$}

\affiliation{ 
$^1$ V. A. Fock Institute of Physics, St. Petersburg
State University, Petrodvorets, Oulianovskaya 1, 198504,
St. Petersburg, Russia
\\
$^2$  Petersburg Nuclear Physics Institute, 188300, Gatchina, St.
Petersburg, Russia}
\begin{abstract}
QED theory of multiphoton cascade transitions in atoms and ions is developed. This theory allows for the accurate description of the process important for astrophysical studies of the cosmological hydrogen recombination. In particular the $ 3s\rightarrow1s+2\gamma $, $ 4s\rightarrow1s+2\gamma $ and $ 3p\rightarrow1s+3\gamma $ processes are considered and some controversies existing in the literature are resolved.
\end{abstract}
\maketitle

\section{Introduction}

The interest to the multiphoton cascade transitions in hydrogen during the last decade was triggered by the accurate measurements of the asymmetry in the temperature and polarization distribution of the Cosmic Microwave Background (CMB) \cite{1}, \cite{2}. The launching of the Planck Surveour enables to perform the measurements with accuracy $ 0.1\% $. It is a challenge to the theory to perform the calculations of the properties of CMB with the same accuracy. For this purpose the adequate theory of the cosmological hydrogen recombination should be developed. The modern theory of this recombination starts from the works by Zel'dovich, Kurt and Sunyaev \cite{3} and by Peebles \cite{4}. According to \cite{3}, \cite{4} the one-photon transitions from the upper levels to the lower ones did not permit the hydrogen atom to recombine, i.e. to reach the ground state. Each photon released in the one-photon transition in atom was immediately absorbed by another atom. This reabsorption process did not allow the radiations to escape the interaction with the matter. However if the atom arrives in the $ 2s $-state, then it decays via the two-photon transition. These two photons escape the reabsorption and the recombination occurs. It was first established in \cite{3}, \cite{4} where $ 2s-1s $ transition was found to be the main channel for the radiation escape and formation of CMB. Hence the recent properties of the CMB are essentially defined by the two-photon processes during the cosmological recombination epoch.

Apart from $ 2s-1s $ transition, as it was noted recently in \cite{5} the two-photon decays from the excited states with the principal quantum number $ n>2 $ also can contribute at the $ 1\% $ level of accuracy. This idea was further developed and intensively discussed in \cite{6}-\cite{8}. There is a difference between the decay of $ ns $ $ (n>2) $, $ nd $ states and the decay of $ 2s $ state. This difference is due to the presence of cascade transitions as the dominant decay channels in case of $ ns $ $ (n>2) $, $ nd $ levels. For the $ 2s $ level the cascades are absent. The cascade photons can be effectively reabsorbed and therefore the problem of separation of the "pure" two-photon emission from the cascade photons arises in connection with the escape probability.

A problem of cascade separation appeared to be nontrivial and caused several controversies in the literature. For the first time this question was raised in \cite{9} for the two-photon transitions in the two electron Highly Charged Ions (HCI). The same problem was considered later in \cite{10}. In \cite{11}, \cite{12} a general QED approach was developed which allowed for the description of the few-photon cascade transitions. This approach was based on the F. Low theory of the line profile in QED \cite{13}. The new interest to cascade separation problem did arise in the context of the cosmological hydrogen recombination. The ambiguity of this separation was shown  in \cite{14} where it was demonstrated that the contributions of the cascade, "pure" two-photon and interference terms can vary depending on the method of calculation and only the total transition rate remains invariant. 

The controversies in the cascade description appeared in connection with the cascade regularization methods. It is convenient to describe these controversies using the simplest example: the two-photon transitions $ 3s\rightarrow1s+2\gamma $. In hydrogen we can restrict ourselves with the nonrelativistic theory and only the electric dipole ($ E1 $) transitions. In $ 3s\rightarrow 1s+2\gamma $ transition a single cascade $ 3s\rightarrow2p+\gamma\rightarrow1s+2\gamma $ should be taken into account.

A total transition rate $ W^{2\gamma}_{3s-1s} $ can be written as
\begin{eqnarray}
\label{1}
W^{2\gamma}_{3s-1s}=\frac{1}{2}\int\limits^{\omega_0}_0dW^{2\gamma}_{3s-1s}(\omega)\;,
\end{eqnarray}
where $ dW^{2\gamma}_{3s-1s}(\omega) $ is the differential transition rate, $ \omega $ is the frequency of one of the emitted photons, $ \omega_0=E_{3s}-E_{1s} $. The differential transition rate $dW^{2\gamma}_{3s-1s}(\omega)$ consists of three terms: cascade contribution, "pure" two-photon contribution and the interference contribution:
\begin{eqnarray}
\label{2}
dW^{2\gamma}_{3s-1s}=dW^{2\gamma(cascade)}_{3s-1s}+dW^{2\gamma(pure)}_{3s-1s}+dW^{2\gamma(interference)}_{3s-1s}\;.
\end{eqnarray}
The cascade contribution can be presented as the sum of the contributions of two cascade links (resonances):
\begin{eqnarray}
\label{3}
dW^{2\gamma(cascade)}=dW^{2\gamma(resonance\;1)}_{3s-2p-1s}+dW^{2\gamma(resonance\;2)}_{3s-2p-1s}\;,
\end{eqnarray}
where two resonant frequencies are: $ \omega^{res1}=E_{3s}-E_{2p} $ and $ \omega^{res2}=E_{2p}-E_{1s}$. The corresponding resonance contributions were presented in \cite{14} on the basis of the QED approach developed in \cite{11}, \cite{12}. These contributions look like
\begin{eqnarray}
\label{4}
dW^{2\gamma(resonance\;1)}_{3s-2p-1s}=\frac{\Gamma_{3s}+\Gamma_{2p}}{\Gamma_{2p}}\frac{W^{1\gamma}_{3s-2p}(\omega^{res1})W^{1\gamma}_{2p-1s}(\omega^{res2})d\omega}{(\omega-\omega^{res1})^2+\frac{1}{4}(\Gamma_{3s}+\Gamma_{2p})^2}\;,
\end{eqnarray}

\begin{eqnarray}
\label{5}
dW^{2\gamma(resonance\;2)}_{3s-2p-1s}=\frac{W^{1\gamma}_{3s-2p}(\omega^{res1})W^{1\gamma}_{2p-1s}(\omega^{res2})d\omega}{(\omega-\omega^{res2})^2+\frac{1}{4}\Gamma_{2p}^2}\;.
\end{eqnarray}

Note that the factor $ \frac{\Gamma_{3s}+\Gamma_{2p}}{\Gamma_{2p}} $ in Eq. (\ref{4}) was lost in \cite{14} which led to the wrong values for $ W^{2\gamma}_{3s-1s} $, $ W^{2\gamma}_{3d-1s}$ transition rates different from correct ones in the third digit in case of $ 3s $. This mistake was noticed in \cite{15}.

Here $ \Gamma_{3s} $, $ \Gamma_{2p} $ are the total widths of the levels $ 3s $, $ 2p $ and $ W^{1\gamma}_{3s-2p} $, $ W^{1\gamma}_{2p-1s} $ are the one-photon transition rates. In the nonrelativistic limit $ \Gamma_{3s}=W^{1\gamma}_{3s-2p} $,  $ \Gamma_{2p}=W^{1\gamma}_{2p-1s}$. Then, integrating Eqs. (\ref{4}), (\ref{5}) over $ \omega $ and taking into account Eq. (\ref{1}) we find

\begin{eqnarray}
\label{6}
\frac{1}{2}\int\limits^{\omega_0}_0dW^{2\gamma(resonance\;1)}_{3s-2p-1s}=\frac{1}{2}W^{1\gamma}_{3s-2p}=\frac{1}{2}\Gamma_{3s}\;,
\end{eqnarray}

\begin{eqnarray}
\label{7}
\frac{1}{2}\int\limits^{\omega_0}_0dW^{2\gamma(resonance\;2)}_{3s-2p-1s}=\frac{1}{2}W^{1\gamma}_{3s-2p}=\frac{1}{2}\Gamma_{3s}\;.
\end{eqnarray}
Hence,
\begin{eqnarray}
\label{7a}
 W^{2\gamma(cascade)}_{3s-1s}=\Gamma_{3s} 
\end{eqnarray}
and
\begin{eqnarray}
\label{8}
W^{2\gamma}_{3s-1s}=\Gamma_{3s}+\frac{1}{2}\int\limits^{\omega_0}_0[dW^{2\gamma(pure)}_{3s-1s}+dW^{2\gamma(interference)}_{3s-1s}]\;.
\end{eqnarray}
From Eq. (\ref{8}) follows that the deviation of $ W^{2\gamma}_{3s-1s} $ from $ \Gamma_{3s} $ is quite small; actually this deviation arrives only in the 5th digit.

A controversy did arise in connection with the regularization of the divergent cascade terms in the integral (1). In \cite{6}, \cite{7}, \cite{8} as in some further papers the regularization was performed by introducing the widths for the intermediate $ np $ states, i.e. replacing the energy $ E_{np} $ by $ E_{np}-\frac{i}{2}\Gamma_{np} $. This replacement was made phenomenologically within the Quantum Mechanical (QM) description. In case of $ 3s-2p-1s $ cascade $ \Gamma_{np}=\Gamma_{2p} $. Therefore the contribution of the resonance 1 instead of Eq. (\ref{4}) looked like
\begin{eqnarray}
\label{9}
dW^{2\gamma(resonance\;1)}_{3s-2p-1s}=\frac{W^{1\gamma}_{3s-2p}(\omega^{res1})W^{1\gamma}_{2p-1s}(\omega^{res2})d\omega}{(\omega-\omega^{res1})^2+\frac{1}{4}\Gamma_{2p}^2}\;,
\end{eqnarray}
while the contribution of $ dW^{2\gamma(resonance\;2)}_{3s-2p-1s} $ remained the same as in Eq. (\ref{5}). In principle. Eq. (\ref{9}) for the cascade transition $ 3s-2p-1s $ can be considered as an approximation to Eq. (\ref{4}) since
\begin{eqnarray}
\label{10}
 \frac{\Gamma_{3s}+\Gamma_{2p}}{\Gamma_{2p}}=\Gamma_{2p}(1+\frac{\Gamma_{3s}}{\Gamma_{2p}})\simeq\Gamma_{2p}(1+0.01)\;.
\end{eqnarray}
However, S.G. Karshenboim, V.G. Ivanov and J. Chluba \cite{15} insisted that the correct expression for the contribution of the first resonance to the $ W^{2\gamma}_{3s-1s} $ transition rate is Eq. (\ref{9}) but not Eq. (\ref{4}). 

In the present paper we demonstrate explicitly that our expression (\ref{4}) is exact and the expression (\ref{9}) can be considered only as an approximation. Moreover we prove that the statement made by S.G. Karshenboim et al in \cite{15} concerning the independence of the two-photon transition rates on the initial state widths, is definitely wrong.

Note that we do not deny, in principle, the usefulness of QM approach; still this approach should be applied with more caution. All these circumstances require full clarification which will be given in the present paper. Completing the introduction we have to stress that the insertion of Eq. (\ref{9}) in the integral Eq. (\ref{1}) gives exactly the same result Eq. (\ref{6}) as the insertion  of Eq. (\ref{4}). One could think that both methods of regularizations, the QED one applied in \cite{11}, \cite{12}, \cite{14}, \cite{16}, \cite{17} and the QM one applied in \cite{6}, \cite{7}, \cite{8}, \cite{15}, are equivalent. However, this is not the case for three reasons. First, this equivalence for the cascade contributions is approximate. Eqs. (\ref{6}), (\ref{7}) are valid up to the small corrections of the order $ \Gamma_{3s}/\omega_{0} $. With the same accuracy holds the mentioned equivalence. The statement that the cascades do not contribute at all to the radiation escape is also approximate since the cascades can not be separated exactly from the "pure two-photon" contribution. Second, the interference contribution also requires regularization and depends on the regularization method. The integral contributions from the interference terms, unlike the integral cascade contributions are not equivalent for the different regularization schemes. Third, in the astrophysical applications the frequency distributions for the two-photon decays are converted usually with some other functions. This also violates the equivalence mentioned above. The total contribution of all the excited states to the radiation escape according to \cite{6} is about $ 0.4 \% $. Thus the error due to the employment of the wrong regularization scheme hardly can exceed $ 0.1\% $. Nevertheless having in mind rapidly growing accuracy of the astrophysical measurements of the CMB the development of the accurate QED theory of the processes in hydrogen, connected with the cosmological recombination seems to be necessary.

The paper is organized as follows. In section II we start with the QED derivation of the Lorentz profile for the one-photon transition from the excited state to the ground state. This derivation repeats shortly the derivations in \cite{11}, \cite{12} but is necessary to introduce the basic formula and notations. As an example the Lyman-alpha transition $ 2p-1s $ is considered. In section III the two-photon transition rate to the ground state from the $ ns $-state in the presence of cascades is described in general. In section IV the regularization of the two-photon transition $ 3s-1s $ is analysed. For this transition the QED regularization scheme \cite{11}, \cite{12} deviates from the QM approach employed in \cite{6}-\cite{8}, \cite{15}. In section V the "pure" two-photon and interference contribution to the $ 3s-1s$  two photon transition rate are described. In section VI the same derivations are made for the two-photon $ 4s-1s $ transition: there is an important difference between $ 4s-1s $ and $ 3s-1s $ two-photon transitions due to the existence of several cascade channels in case of $ 4s-1s $. In section VII the 3-photon transitions are analysed with 3-photon decay $ 3p-1s $ as an example. Concluding remarks are presented in section VIII. 
\section{One-photon transition to the ground state}

The full QED description of any process in an atom should start with the ground state and end up with the ground state too, i.e. the excitation of the decaying state should be always included. For the resonant processes, e.g. for the resonant photon scattering the absorption part of the process can be well separated from the emission part, so that the description of the decay process independent on the excitation becomes possible. In this way the theory of the multiphoton processes in atoms was developed in  \cite{11}, \cite{12}. A simplified version of the theory which starts directly from the excited state was considered in \cite{16}. This version allows for the correct description of the complicated multiphoton processes with cascades but does not allow to trace down the details of the regularization of the divergent cascade contributions, i.e. one has to refer to the more elaborate evaluations  \cite{11}, \cite{12}. Since the controversies mentioned above in section I, concern namely the regularization methods, in this work we follow the description formulated in  \cite{11}, \cite{12}.

Having in mind the recombination processes in hydrogen atom we consider first the resonance photon scattering on the ground $ 1s $ state with resonances corresponding to the $ np $ states. In our derivations we will fully neglect the photons other than $ E1 $ which is reasonable for the neutral hydrogen. It is important to stress that we consider the free atoms which are excited by the photons released by the source which line widths is comparable (or larger) then the natural line widths of the resonance atomic state. Thus we exclude the special cases of the excitation by the laser with the narrow bandwidths or something equivalent. Our condition (broad source width) should correspond the cosmological recombination situation when every atom is excited by the photons emitted by another atom. The Feynman graph corresponding to the resonant photon scattering is depicted in Fig. 1a.

The $ S $-matrix element, corresponding to Fig. 1a, i.e. second-order scattering process, looks like
\begin{eqnarray}
\label{11}
S_{1s}^{(2)sc}=(-ie)^2\int d^4x_1d^4x_2\overline{\psi}_{1s}(x_1)\gamma_{\mu_1}A_{\mu_1}^{*(\vec{k}_f\vec{e}_f)}(x_1)S(x_1,x_2)\gamma_{\mu_2}A_{\mu_2}^{(\vec{k}_i\vec{e}_i)}(x_2)\psi_{1s}(x_2)\;,
\end{eqnarray}
where
\begin{eqnarray}
\label{12}
\psi_A(x)=\psi_A(\vec{r})e^{-iE_At}\;,
\end{eqnarray}
$ \psi_A(\vec{r}) $ is the solution of the Dirac equation for the atomic electron, $ E_A $ is the Dirac energy, $ \overline{\psi}_A=\psi^+_A\gamma_0 $ is the Dirac conjugated wave function, $ \gamma_{\mu}\equiv(\gamma_0,\vec{\gamma})$ are the Dirac matrices and $ x\equiv(\vec{r},t) $ is the space-time coordinate. The photon field or the photon wave function $ A_{\mu}(x) $ looks like
\begin{eqnarray}
\label{13}
A_{\mu}^{(\vec{k},\vec{e})}=\sqrt{\frac{2\pi}{\omega}}e_{\mu}e^{i(\vec{k}\vec{r}-\omega t)}=\sqrt{\frac{2\pi}{\omega}}e^{-i\omega t}A_{\mu}^{(\vec{k},\vec{e})}(\vec{r})\;,
\end{eqnarray}
where $ e_{\mu} $ are the components of the photon polarization four-vector ($ \vec{e} $ is 3-dimensional polarization vector for real photons), $ k\equiv(\vec{k},\omega) $ is the photon momentum four-vector, $ \vec{k} $ is the wave vector, $ \omega=|\vec{k}| $ is the photon frequency. Eq. (\ref{13}) corresponds to the absorbed photon and  $ A^{*(\vec{k},\vec{e})}_{\mu} $ corresponds to the emitted photon. Finally, the electron propagator for the bound electron it is convenient to present in the form of the eigenmode decomposition with respect to one-electron eigenstates \cite{18}
\begin{eqnarray}
\label{14}
S(x_1,x_2)=\frac{1}{2\pi i}\int\limits^{\infty}_{-\infty}d\omega e^{-i\omega(t_1-t_2)}\sum\limits_{n}\frac{\psi_n(\vec{r}_1)\overline{\psi}_n(\vec{r}_2)}{\omega-E_n(1-i0)}\;.
\end{eqnarray}
Insertion of the expressions (\ref{12})-(\ref{14}) into Eq. (\ref{11}) and integration over time and frequency variables leads to
\begin{eqnarray}
\label{15}
S_{1s}^{(2)sc}=-2\pi i\delta(\omega_f-\omega_i)e^2\sum\limits_{n}\frac{(\gamma_{\mu}A^{*(\vec{k}_f,\vec{e}_f)}_{\mu})_{1sn}(\gamma_{\mu}A^{(\vec{k}_i,\vec{e}_i)}_{\mu})_{n1s}}{\omega_f+E_{1s}-E_n}\;.
\end{eqnarray}
The amplitude $ U $ of the elastic photon scattering is related to the $ S $-matrix element via \cite{18}
\begin{eqnarray}
\label{16}
S=-2\pi i\delta(\omega_f-\omega_i)U\;.
\end{eqnarray}
Accordingly, we will obtain the scattering amplitude 
\begin{eqnarray}
\label{17}
U^{(2)sc}_{1s}=e^2\sum\limits_{n}\frac{(\gamma_{\mu}A^{*(\vec{k}_f,\vec{e}_f)}_{\mu})_{1sn}(\gamma_{\mu}A^{(\vec{k}_i,\vec{e}_i)}_{\mu})_{n1s}}{\omega_f+E_{1s}-E_n}\;,
\end{eqnarray}
where the energy conservation law implies that $ |\vec{k_f}|=|\vec{k_i}| $.

For the resonant scattering process the photon frequency $ \omega_i=\omega_f $ is close to the energy difference between two atomic levels. In case of $ np $ resonance $ \omega_i\simeq E_{np}-E_{1s} $. Accordingly we have to retain only one term in the sum over $ n $ in Eq. (\ref{17})
\begin{eqnarray}
\label{18}
U^{(2)sc}_{1s(np)}=e^2\frac{(\gamma_{\mu}A^{*(\vec{k}_f,\vec{e}_f)}_{\mu})_{1snp}(\gamma_{\mu}A^{(\vec{k}_i,\vec{e}_i)}_{\mu})_{np1s}}{\omega_f+E_{1s}-E_{np}}\;.
\end{eqnarray}

Eq. (\ref{18}) reveals that in the resonance approximation the scattering amplitude is factorized into an emission and absorption parts. The energy denominator should be attached to the emission or absorption part depending on what we want to describe: emission or absorption process. In particular, the first-order emission amplitude can be expressed as
\begin{eqnarray}
\label{19}
U^{em}_{np1s}=e\frac{(\gamma_{\mu}A^{*(\vec{k}_f,\vec{e}_f)}_{\mu})_{1snp}}{\omega_f+E_{1s}-E_{np}}\;.
\end{eqnarray}

The nonresonant corrections to the resonance approximation, first introduced  in \cite{13} were recently investigated in \cite{19}-\cite{21}. The role of these corrections appeared to be negligible in most cases. These corrections arise when one takes into account the terms other than the resonant one in sum over $ n $ in Eq. (\ref{17}). The same concerns nonresonant contribution to the scattering amplitude which arises when we interchange the position of the photon lines in Fig. 1, i.e. when the emission of the photon occurs prior to the absorption.

The energy conservation law which follows from Eq. (\ref{16}) reads
\begin{eqnarray}
\label{20a}
\omega_i=\omega_f\;.
\end{eqnarray}
The resonance condition one can write in the form:
\begin{eqnarray}
\label{20b}
|\omega_i-E_{np}+E_{1s}|=|\omega_f-E_{np}+E_{1s}|\leqslant \Gamma_{np}\;.
\end{eqnarray}
In cases, when we can neglect $ \Gamma_{np} $ in Eq. (\ref{20b}) this equation takes the form of the energy conservations law
\begin{eqnarray}
\label{20c}
\omega_f=E_{np}-E_{1s}\;.
\end{eqnarray}
In particular we can use Eq.(\ref{20c}) in the numerator of Eq. (\ref{19}) but not in its denominator.

To derive the Lorentz profile for the emission process we follow the Low procedure \cite{13}, i.e. insert infinite number of the self-energy corrections in the resonance approximation into the electron propagator in Fig. 1a. The first term of the corresponding Feynman graph sequence is depicted in Fig. 1b. Employing the photon propagator in the Feynman gauge in the form
\begin{eqnarray}
\label{20}
D_{\mu_1 \mu_2}(x_1-x_2)=\frac{1}{2\pi i}\int\limits^{\infty}_{-\infty}d\Omega I_{\mu_1 \mu_2}(|\Omega|, r_{12})e^{-i\Omega(t_1-t_2)}\;,
\end{eqnarray}
\begin{eqnarray}
\label{21}
I_{\mu_1 \mu_2}=\frac{\delta_{\mu_1 \mu_2}}{r_{12}}e^{i|\Omega| r_{12}}\;,
\end{eqnarray}
where $ x\equiv(\vec{r}, t) $, $ r_{12}=|\vec{r}_1-\vec{r}_2| $ and defining the matrix element of the electron self-energy operator as \cite{22}
\begin{eqnarray}
\label{22}
(\widehat{\Sigma}(\xi))_{AB}=\frac{e^2}{2\pi i}\sum\limits_{n}\int d\Omega\frac{(\gamma_{\mu_1}\gamma_{\mu_2}I_{\mu_1\mu_2}(|\Omega|,r_{12}))_{AnnB}}{\xi-\Omega-E_n(1-i0)}\;,
\end{eqnarray}
 we obtain the following expression for the correction to the scattering amplitude \cite{12}:
\begin{eqnarray}
\label{23}
U^{(4)sc.}_{1s}=e^2\sum\limits_{n_1 n_2}\frac{(\gamma_{\mu}A^{*(\vec{k}_f,\vec{e}_f)}_{\mu})_{1sn_1}(\widehat{\Sigma}(\omega+E_{1s}))_{n_1n_2}(\gamma_{\mu}A^{(\vec{k}_i,\vec{e}_i)}_{\mu})_{n_21s}}{(\omega_f+E_{1s}-E_{n_1})(\omega+E_{1s}-E_{n_2})}\;.
\end{eqnarray}
The resonance approximation implies $ n_1=n_2=np $. Then taking into account Eq. (\ref{18}) we can write
\begin{eqnarray}
\label{24}
U^{(4)sc.}_{1s(np)}=U^{(2)sc.}_{1s(np)}\frac{(\widehat{\Sigma}(\omega_f+E_{1s}))_{np,np}}{\omega_f+E_{1s}-E_{np}}\;.
\end{eqnarray}
Repeating these insertions in the resonance approximation leads to a geometric progression. Summation of this progression yields
\begin{eqnarray}
\label{25}
U^{(4)sc.}_{1s(np)}=e^2\frac{\gamma_{\mu}A^{*(\vec{k}_f,\vec{e}_f)}_{\mu})_{1snp}(\gamma_{\mu}A^{(\vec{k}_i,\vec{e}_i)}_{\mu})_{np1s}}{(\omega_f+E_{1s}-E_{2p}-(\widehat{\Sigma}(\omega_f+E_{1s}))_{np,np}}\;.
\end{eqnarray}
The emission amplitude looks like
\begin{eqnarray}
\label{26}
U^{em}_{np 1s}=e\frac{(\gamma_{\mu}A^{*(\vec{k}_f\vec{e}_f)}_{\mu})_{1snp}}{\omega_f+E_{1s}-E_{np}-(\widehat{\Sigma}(\omega_f+E_{1s}))_{np,np}}\;.
\end{eqnarray}
The operator $ \widehat{\Sigma}(\omega_f+E_{1s}) $ can be expanded around the value $ \omega_f+E_{1s}=E_{np} $
\begin{eqnarray}
\label{27}
\widehat{\Sigma}(\omega_f+E_{1s})=\widehat{\Sigma}(E_{np})+(\omega_f+E_{1s}-E_{np})\widehat{\Sigma '}(E_{np})+...\;,
\end{eqnarray}
where $ \widehat{\Sigma '}(E_{np})\equiv\frac{d}{d\xi}\widehat{\Sigma}(\xi)|_{\xi=E_{np}} $. The first two terms of the expansion (\ref{27}) are ultraviolet divergent and require the renormalization. The methods of the renormolization in the bound electron QED are described, for example in \cite{23}. In order to obtain the line profile for the emission process we retain the first term of the expansion (\ref{27}) and consider the energy denominator in Eq. (\ref{25}) as a complex quantity:
\begin{eqnarray}
\label{28}
(\widehat{\Sigma}(E_{np}))_{np,np}=L^{SE}_{np}-\frac{i}{2}\Gamma_{np}\;.
\end{eqnarray}
Here $ L^{SE}_{np} $ is the electron self-energy contribution to the electron Lamb shift and $ \Gamma_{np} $ is the one-photon radiative level width \cite{22}. Apart from $ L^{SE} $ contribution
there is also the vacuum polarization $ L^{VP} $ contribution \cite{12}, but the vacuum polarization contribution is pure real and does not change the imaginary part in Eq. (\ref{28}). Now the emission amplitude reads
\begin{eqnarray}
\label{29}
U^{em}_{np-1s}=e\frac{(\gamma_{\mu}A^{*(\vec{k}_f\vec{e}_f)}_{\mu})_{1snp}}{\omega_f+E_{1s}-E_{np}-L_{np}+\frac{i}{2}\Gamma_{np}}\;,
\end{eqnarray}
where $ L_{np}=L^{SE}_{np}+L^{VP}_{np} $.

As a next step one has to take the amplitude Eq. (\ref{29}) by square modulus, then integrate over the photon emission directions $ \vec{\nu}_f $ and sum over the photon polarizations. Defining the one-photon transition rate for the transition $ np-1s $ like
\begin{eqnarray}
\label{30}
W^{1\gamma}_{np-1s}=2\pi\omega^2_{res}\sum\limits_{\vec{e}_f}\int \frac{d\vec{\nu_f}}{(2\pi)^3}|(\gamma_{\mu}A^{*(\vec{k}_f,\vec{e}_f)}_{\mu})_{np1s}|^2\;,
\end{eqnarray}
where $ \omega_{res} $ is the resonant photon frequency. In Eq. (\ref{30}) it is assumed also the summation over the degenerate substates of the final state and averaging over the degenerate substates of the initial state. These operations we will not designate explicitly since it does not influence our argumentation. The same will concern the two-photon and three-photon transitions in the subsequent sections.

From Eq. (\ref{29}) we obtain for the absolute probability of the photon emission with the frequency in the interval between $ \omega_f $ and $ \omega_f+d\omega_f $
\begin{eqnarray}
\label{31}
dw_{np-1s}(\omega_f)=\frac{1}{2\pi}\frac{W^{1\gamma}_{np-1s}d\omega_f}{(\omega_f+E_{1s}-E_{np}-L_{np})^2+\frac{1}{4}\Gamma^2_{np}}\;.
\end{eqnarray}
Due to the factor $ \frac{1}{2\pi} $ the Lorentz profile Eq. (\ref{31}) is normalized to unity for the Lyman-alpha transition
\begin{eqnarray}
\label{32}
\int\limits_0^{\infty}dw_{2p-1s}=1\;.
\end{eqnarray}
In case $ n>2 $ the Lorentz profile is normalized to the branching ratio for the transition $ np-1s $:
\begin{eqnarray}
\label{33}
\int\limits_0^{\infty}dw_{np-1s}=\frac{W^{1\gamma}_{np-1s}}{\Gamma_{np}}=b^{1\gamma}_{np-1s}\;.
\end{eqnarray}

The Lamb shift for the ground $ 1s $ state enters the energy denominator in Eq. (\ref{31}) in  different way. Insertions of the electron self-energy corrections in the outer electron lines in Fig. 1, unlike the insertions in the internal electron line lead to the singularities when the intermediate states in propagators are equal to $ 1s $. This singularities are not connected with the frequency resonances. To regularize these singularities one has to introduce Gel-Mann and Low \cite{24} adiabatic $ S $-matrix as it was done in \cite{25}. It was demonstrated that the summation of the infinite series of the singular in the adiabatic parameter $ \lambda $ terms can be converted to the exponential factor. The amplitude Eq. (\ref{25}) should be replaced by
\begin{eqnarray}
\label{34}
\lim\limits_{\lambda\rightarrow 0}U^{sc}_{1s(np)}(\lambda)=e^2\frac{(\gamma_{\mu}A^{*(\vec{k}_f\vec{e}_f)}_{\mu})_{1snp}(\gamma_{\mu}A^{(\vec{k}_i\vec{e}_i)}_{\mu})_{np1s}}{\omega_f+E_{1s}+L_{1s}-E_{np}-L_{np}+\frac{i}{2}\Gamma_{np}}e^{-\frac{i}{\lambda}(\widehat{\Sigma}(E_{1s}))_{1s1s}}\;.
\end{eqnarray}
Since for the ground state the matrix element $ (\widehat{\Sigma}(E_{1s}))_{1s1s} $ is pure real, for the probability this gives
\begin{eqnarray}
\label{35}
\lim\limits_{\lambda\rightarrow 0}|e^{-\frac{i}{\lambda}(\widehat{\Sigma}(E_{1s}))_{1s1s}}|=1
\end{eqnarray}
and thus the Lamb shift $ L_{1s} $ arrives in the expression (\ref{31}) for the Lorentz profile.
Note, however, that if we apply Eq. (\ref{35}) to the excited state and take into account the width of the excited level we will obtain zero transition probability. Strictly speaking this means that it is incorrect to evaluate the transition probabilities via the nondiagonal $ S $-matrix elements as is usually done in QED for atoms and it is necessary to start with the process of excitation using the procedure described in the present paper. However in most cases Eq. (\ref{35}) can be ignored and the correct results for transitions rates are obtained in a standard way, evaluating the square modules of the nondiagonal $ S $-matrix elements. Only in the special situations as in case of the multiphoton cascade transitions considered in the subsequent sections of the present paper, more refined analysis is required.

\section{Two-photon $ns-1s$ transition}

In this section we describe the two-photon transition to the ground state using as an example $ ns-1s $ two-photon transitions. According to our approach we have to start with the Feynman graph depicted in Fig. 2a. The two-photon resonant excitation is the most natural and convenient way to describe the excitation process in this case. The resonance condition is
\begin{eqnarray}
\label{36}
\omega_{i1}+\omega_{i2}=\omega_0^{ns}=E_{ns}-E_{1s}\;.
\end{eqnarray}

Constructing the $ S $-matrix element corresponding to the Feynman graph Fig. 2a, inserting the expressions for the electron and photon wave functions as well as the expressions for the electron propagators Eqs. (\ref{12})-(\ref{14}), integrating over time and frequency variables  and using Eq. (\ref{16}) for the scattering amplitude results
\begin{eqnarray}
\label{37}
U^{(4)sc.}_{1s}=e^4\sum\limits_{n_1n_2n_3}\frac{(\gamma_{\mu_1}A_{\mu_1}^{*(\vec{k}_{f_2}\vec{e}_{f_2})})_{1sn_1}(\gamma_{\mu_2}A_{\mu_2}^{*(\vec{k}_{f_1}\vec{e}_{f_1})})_{n_1n_2}}{(\omega_{f_2}+E_{1s}-E_{n_1})(\omega_{f_2}+\omega_{f_1}+E_{1s}-E_{n_2})}\times\\\nonumber
\frac{(\gamma_{\mu_3}A_{\mu_3}^{(\vec{k}_{i_2}\vec{e}_{i_2})})_{n_2n_3}(\gamma_{\mu_4}A_{\mu_4}^{(\vec{k}_{i_1}\vec{e}_{i_1})})_{n_31s}}{(\omega_{f_2}+\omega_{f_1}-\omega_{i_2}+E_{1s}-E_{n_3})}\;.
\end{eqnarray}
The energy conservation in this process is implemented by the condition
\begin{eqnarray}
\label{38}
\omega_{f_1}+\omega_{f_2}=\omega_{i_1}+\omega_{i_2}
\end{eqnarray}
and the resonance condition is given by Eq. (\ref{36}). From Eq. (\ref{36}) follows the approximate energy conservation law similar to Eq. (\ref{20b})
\begin{eqnarray}
\label{39a}
|\omega_{f_1}+\omega_{f_2}-E_{ns}+E_{1s}|\leqslant \Gamma_{ns}\;,
\end{eqnarray}\
which can be replaced by equation similar to Eq. (\ref{20c})
\begin{eqnarray}
\label{39b}
\omega_{f_1}+\omega_{f_2}=E_{ns}-E_{1s}\;,
\end{eqnarray}
when $ \Gamma_{ns} $ can be neglected.
According to Eqs. (\ref{36}) and (\ref{38}) the last energy denominator in Eq. (\ref{37}) can be replaced by 
\begin{eqnarray}
\label{39}
\omega_{f_2}+\omega_{f_1}-\omega_{i_2}+E_{1s}-E_{n_3}=\omega_{i_1}+E_{1s}-E_{n_3}\;,
\end{eqnarray}
i.e. does not depend on the frequencies of emitted photons.

In the resonance approximation we retain only one term $ n_2=ns $ in the sum over $ n_2 $ which yields

\begin{eqnarray}
\label{40}
U^{(4)sc}_{1s(ns)}=e^4\sum\limits_{n_1}\frac{(\gamma_{\mu_1}A_{\mu_1}^{*(\vec{k}_{f_2}\vec{e}_{f_2})})_{1sn_1}(\gamma_{\mu_2}A_{\mu_2}^{*(\vec{k}_{f_1}\vec{e}_{f_1})})_{n_1ns}}{(\omega_{f_2}+E_{1s}-E_{n_1})(\omega_{f_2}+\omega_{f_1}+E_{1s}-E_{ns})}\sum\limits_{n_3}\frac{(\gamma_{\mu_3}A_{\mu_3}^{(\vec{k}_{i_2}\vec{e}_{i_2})})_{nsn_3}(\gamma_{\mu_4}A_{\mu_4}^{(\vec{k}_{i_1}\vec{e}_{i_1})})_{n_31s}}{(\omega_{i_1}+E_{1s}-E_{n_3})}\;.
\end{eqnarray}
Starting from Eq. (\ref{40}) we can write down the expression for the two-photon emission amplitude as
\begin{eqnarray}
\label{41}
U^{(2)em}_{ns-1s}=e^2\sum\limits_{n_1}\frac{(\gamma_{\mu_1}A_{\mu_1}^{*(\vec{k}_{f_2}\vec{e}_{f_2})})_{1sn1}(\gamma_{\mu_2}A_{\mu_2}^{*(\vec{k}_{f_1}\vec{e}_{f_1})})_{n_1ns}}{(\omega_{f_2}+E_{1s}-E_{n_1})(\omega_{f_2}+\omega_{f_1}+E_{1s}-E_{ns})}
\end{eqnarray}
with the condition Eq. (\ref{36}) remaining valid for the frequencies $\omega_{f_1}, \omega_{f_2}$ due to Eq. (\ref{38}).

Up to now all formulas above in this section were valid for any $ ns $ levels, beginning from $ n=2s $. Now we have to take into account the form of the resonance produced by the second energy denominator in Eq. (\ref{41}). The width of this resonance for $ 2s $ level is defined by the two-photon transition $ 2s\rightarrow 1s+2\gamma $. This width should arrive as the imaginary part of the matrix element of the second-order electron self-energy operator, i.e. from two-loop insertions to the Feynman graph Fig. 2a. The rigorous QED derivation of the two-photon widths from the contributions of the  two-loop Feynman graphs is still absent but the contribution of the $ 2s $ level to the CMB history is very well known from \cite{3}, \cite{4} and later works.

Therefore, we will restrict our studies with $ n>2 $. For $ n>2 $ there is always leading one-photon contribution to the total width $ \Gamma_{ns} $, for example the $ W^{1\gamma}_{3s-2p} $ transition rate in case $ n=3 $. Assuming the existence of such a contribution we will continue our studies by inserting the one-loop electron self-energy corrections to the central propagator in Fig. 1a (the Low procedure). The first term of the Low sequence is depicted in Fig. 2b.

Returning back to the scattering amplitude Eq. (\ref{39}) and proceeding along the same way as in the case of the one-photon decay we obtain an expression similar to Eq. (\ref{23}) in the one-photon case:
\begin{eqnarray}
\label{42}
U^{(6)sc}_{1s(ns)}=e^4\sum\limits_{n_1n_2n_3}\frac{(\gamma_{\mu_1}A_{\mu_1}^{*(\vec{k}_{f_2}\vec{e}_{f_2})})_{1sn_1}(\gamma_{\mu_2}A_{\mu_2}^{*(\vec{k}_{f_1}\vec{e}_{f_1})})_{n_1n_2}}{(\omega_2+E_{1s}-E_{n_1})(\omega_{f_2}+\omega_{f_1}+E_{1s}-E_{n_2})}\times\\\nonumber
\frac{(\widehat{\Sigma}(\omega_{f_1}+\omega_{f_2}+E_{1s}))_{n_2n_3}}{(\omega_{f_2}+\omega_{f_1}+E_{1s}-E_{n_3})}\times\frac{(\gamma_{\mu_3}A_{\mu_3}^{(\vec{k}_{i_2}\vec{e}_{i_2})})_{n_3n_3}(\gamma_{\mu_4}A_{\mu_4}^{(\vec{k}_{i_1}\vec{e}_{i_1})})_{n_41s}}{(\omega_1+E_{1s}-E_{n_4})}\;.
\end{eqnarray}
In the resonance approximation setting $ n_2=n_3=ns $ we have
\begin{eqnarray}
\label{43}
U^{(6)sc}_{1s(ns)}=U^{(4)sc}_{1s(ns)}\frac{(\widehat{\Sigma}(\omega_{f_1}+\omega_{f_2}+E_{1s}))_{nsns}}{(\omega_{f_2}+\omega_{f_1}+E_{1s}-E_{ns})}\;.
\end{eqnarray}
Producing further the Low sequence and performing the summation the arising geometric progression results
\begin{eqnarray}
\label{44}
U^{sc}_{1s(ns)}=e^4\sum\limits_{n_1}\frac{(\gamma_{\mu_1}A_{\mu_1}^{*(\vec{k}_{f_2}\vec{e}_{f_2})})_{n_1ns}(\gamma_{\mu_2}A_{\mu_2}^{*(\vec{k}_{f_1}\vec{e}_{f_1})})_{n_1n_2}}{(\omega_{f_2}+E_{1s}-E_{n_1})}\times\\\nonumber
\frac{1}{\omega_{f_2}+\omega_{f_1}+E_{1s}-E_{ns}-(\widehat{\Sigma}(\omega_{f_1}+\omega_{f_2}+E_{1s}))_{ns,ns}}\sum\limits_{n_2}\frac{(\gamma_{\mu_3}A_{\mu_3}^{(\vec{k}_{i_2}\vec{e}_{i_2})})_{n_1ns}(\gamma_{\mu_4}A_{\mu_4}^{(\vec{k}_{i_1}\vec{e}_{i_1})})_{n_1n_2}}{(\omega_{i_1}+E_{1s}-E_{n_2})}\;.
\end{eqnarray}
The emission amplitude for the two-photon decay process $ ns\rightarrow 1s+2\gamma $ looks like
\begin{eqnarray}
\label{45}
U^{em}_{ns-1s}=e^2\sum\limits_{n_1}\frac{(\gamma_{\mu_1}A_{\mu_1}^{*(\vec{k}_{f_2}\vec{e}_{f_2})})_{n_1ns}(\gamma_{\mu_2}A_{\mu_2}^{*(\vec{k}_{f_1}\vec{e}_{f_1})})_{n_1n_2}}{(\omega_{f_2}+E_{1s}-E_{n_1})}\times\\\nonumber
\frac{1}{\omega_{f_2}+\omega_{f_1}+E_{1s}-E_{ns}-(\widehat{\Sigma}(\omega_{f_1}+\omega_{f_2}+E_{1s}))_{ns,ns}}\;.
\end{eqnarray}
At the point of the resonance we expand the operator 
\begin{eqnarray}
\label{46}
\widehat{\Sigma}(\omega_{f_1}+\omega_{f_2}+E_{1s})=\widehat{\Sigma}(E_{ns})+...
\end{eqnarray}
and using the equality
\begin{eqnarray}
\label{47}
(\widehat{\Sigma}(E_{ns}))_{ns,ns}=L^{SE}_{ns}-\frac{i}{2}\Gamma_{ns}\;,
\end{eqnarray}
arrive at
\begin{eqnarray}
\label{48}
U^{em}_{ns-1s}=e^2\sum\limits_{n_1}\frac{(\gamma_{\mu_1}A_{\mu_1}^{*(\vec{k}_{f_2}\vec{e}_{f_2})})_{n_1ns}(\gamma_{\mu_2}A_{\mu_2}^{*(\vec{k}_{f_1}\vec{e}_{f_1})})_{n_1n_2}}{(\omega_{f_2}+E_{1s}-E_{n_1})(\omega_{f_2}+\omega_{f_1}+E_{1s}-E_{ns}+\frac{i}{2}\Gamma_{ns})}\;.
\end{eqnarray}

In Eq. (\ref{48}) we have omitted the Lamb shift of the $ ns $ level in the second energy denominator. In what follows the Lamb shift will play no significant role in our derivations.

The value $ \Gamma_{ns} $ is defined in a different way for the different $ ns $ states. For example, for $ n=3 $  $\Gamma_{3s}=W^{1\gamma}_{3s-2p}$ since there are no other one-photon decay channels for $ 3s $ level. The further investigations of the two-photon transition probabilities should be performed separately for different $ n $. In the next section we will continue these investigations for $ 3s\rightarrow1s+2\gamma $ transition.
\section{Two-photon $3s-1s$ transition}

The further studies of the $ 3s-1s $ transition we can start with the expression for the emission amplitude Eq. (\ref{48}) written for the case $ ns=3s $. The Feynman graphs for the resonance two-photon scattering with the excitation of $ 3s $ level are depicted in Fig. 3. To the expression Eq. (\ref{48}) we have to add also another term corresponding to the Feynman graph Fig. 3a with the interchanged positions of the $ \vec{k}_{f_1},\vec{e}_{f_1} $ and $ \vec{k}_{f_2},\vec{e}_{f_2} $ photons. This yields
\begin{eqnarray}
\label{49}
U^{em}_{3s-1s}=e^2\sum\limits_{n_1}\left\lbrace \frac{(\gamma_{\mu_1}A_{\mu_1}^{*(\vec{k}_{f_2}\vec{e}_{f_2})})_{1sn_1}(\gamma_{\mu_2}A_{\mu_2}^{*(\vec{k}_{f_1}\vec{e}_{f_1})})_{n_13s}}{\omega_{f_2}+E_{1s}-E_{n_1}}+\frac{(\gamma_{\mu_1}A_{\mu_1}^{*(\vec{k}_{f_1}\vec{e}_{f_1})})_{1sn_1}(\gamma_{\mu_2}A_{\mu_2}^{*(\vec{k}_{f_2}\vec{e}_{f_2})})_{n_13s}}{\omega_{f_1}+E_{1s}-E_{n_1}}\right\rbrace\times\\\nonumber\times \frac{1}{\omega_{f_2}+\omega_{f_1}+E_{1s}-E_{3s}+\frac{i}{2}\Gamma_{3s}}\;.
\end{eqnarray}

For the $ 3s-1s $ two-photon transition only one cascade is possible: $ 3s-2p-1s $. Accordingly, the two new resonance conditions arise (these resonances were defined also in section I):
\begin{eqnarray}
\label{50}
\omega^{res.1}=E_{3s}-E_{2p}\;,
\end{eqnarray}
\begin{eqnarray}
\label{51}
\omega^{res.2}=E_{2p}-E_{1s}\;.
\end{eqnarray}

Consider first cascade contribution to Eq. (\ref{49}). For this purpose we have to set $ n_1=2p $. Then
\begin{eqnarray}
\label{52}
U^{em,\; cascade}_{3s-2p-1s}=e^2\left\lbrace \frac{(\gamma_{\mu_1}A_{\mu_1}^{*(\vec{k}_{f_2}\vec{e}_{f_2})})_{1s2p}(\gamma_{\mu_2}A_{\mu_2}^{*(\vec{k}_{f_1}\vec{e}_{f_1})})_{2p3s}}{\omega_{f_2}+E_{1s}-E_{2p}}+\frac{(\gamma_{\mu_1}A_{\mu_1}^{*(\vec{k}_{f_1}\vec{e}_{f_1})})_{1s2p}(\gamma_{\mu_2}A_{\mu_2}^{*(\vec{k}_{f_2}\vec{e}_{f_2})})_{2p3s}}{\omega_{f_1}+E_{1s}-E_{2p}}\right\rbrace\times\\\nonumber\times \frac{1}{\omega_{f_2}+\omega_{f_1}+E_{1s}-E_{3s}+\frac{i}{2}\Gamma_{3s}}\;.
\end{eqnarray}

The first term in the curly  brackets describes the resonance (\ref{50}), the second term describes the resonance (\ref{51}) (see Appendix A). Applying the Low procedure (insertions and summation of the infinite chain of the electron self-energy corrections) to the upper electron propagators in Fig. 3b we find
\begin{eqnarray}
\label{53}
U^{em,\; cascade}_{3s-2p-1s}=e^2\left\lbrace \frac{(\gamma_{\mu_1}A_{\mu_1}^{*(\vec{k}_{f_2}\vec{e}_{f_2})})_{1sn_1}(\gamma_{\mu_2}A_{\mu_2}^{*(\vec{k}_{f_1}\vec{e}_{f_1})})_{2p3s}}{\omega_{f_2}+E_{1s}-E_{2p}+\frac{i}{2}\Gamma_{2p}}+\frac{(\gamma_{\mu_1}A_{\mu_1}^{*(\vec{k}_{f_1}\vec{e}_{f_1})})_{1sn_1}(\gamma_{\mu_2}A_{\mu_2}^{*(\vec{k}_{f_2}\vec{e}_{f_2})})_{2p3s}}{\omega_{f_1}+E_{1s}-E_{2p}+\frac{i}{2}\Gamma_{2p}}\right\rbrace \times\\\nonumber\times\frac{1}{\omega_{f_2}+\omega_{f_1}+E_{1s}-E_{3s}+\frac{i}{2}\Gamma_{3s}}\;.
\end{eqnarray}

Now we take $ U^{em, cascade}_{3s-2p-1s} $ by square modulus, integrate over the emitted photons directions and sum over the polarizations of both photons. The formula (\ref{30}) should be used for presentation of the results of these integrations and summation via the one-photon transition rates. Consider first the square modulus of the first term in the curly brackets and the factor outside the curly brackets in Eq. (\ref{53}). This term is represented by Fig. 3a and corresponds to the contribution of the resonance 1 in Eq. (\ref{50}). In this case we are interested to derive the Lorentz line profile for the upper link of the cascade $ 3s-2p-1s $. Therefore we have to integrate first over frequency of the second emitted photon, i.e. $ \omega_{f_2} $. In principle the integration over both photon frequencies should be done with Eq. (\ref{39a}) taken into account, i.e.
\begin{eqnarray}
\label{54a}
\int\limits_0^{\omega_{max}}d\omega_{f_1}\int\limits_0^{\omega_{1}}d\omega_{f_2}=\frac{1}{2}\int\limits_0^{\omega_{max}}d\omega_{f_1}\int\limits_0^{\omega_{max}}d\omega_{f_2}\;,
\end{eqnarray}
where $ \omega_{max}=E_{2s}-E_{1s} $.

Eq. (\ref{54a}) holds due to the symmetry of Eq. (\ref{53}) with respect to permutation $ \omega_{f_1}\leftrightarrows \omega_{f_2} $.

The integration over the frequency $ \omega_{f_2} $ in Eq. (\ref{53}) we perform in the complex plane. Since only the pole terms contribute we can extend the interval of integration to $ (-\infty, +\infty) $ and not to refer to Eq. (\ref{39a}) or (\ref{54a}). Then using Cauchy theorem after some algebraic transformation (see for details the Appendix A) we obtain the cascade contribution (resonance 1) to the differential branching ratio
\begin{eqnarray}
\label{54}
db^{2\gamma(resonance\;1)}_{3s-2p-1s}(\omega)=\frac{1}{2\pi}\frac{\Gamma_{3s}+\Gamma_{2p}}{\Gamma_{3s}\Gamma_{2p}}\frac{W_{3s-2p}^{1\gamma}(\omega^{res. 1})W_{2p-1s}^{1\gamma}(\omega^{res. 2})d\omega}{(\omega-\omega^{res. 1})^2+\frac{1}{4}(\Gamma_{3s}+\Gamma_{2p})^2}\;
\end{eqnarray}
(here we have changed the notation for the frequency from  $ \omega_{f_1} $ to $ \omega $).

The differential branching ratio $ db^{2\gamma} $ is connected with the differential transition rate $ dw^{2\gamma}_{ns-1s}(\omega) $ via
\begin{eqnarray}
\label{55}
db^{2\gamma}_{ns-1s}(\omega)=\frac{dw^{2\gamma}_{ns-1s}}{\Gamma_{ns}}\;.
\end{eqnarray}

This definition concerns not only the cascade contributions but all the contributions in Eq. (\ref{2}) for the two-photon decay of any $ ns $-state:
\begin{eqnarray}
\label{56}
db^{2\gamma}_{ns-1s}=db^{2\gamma(cascade)}_{ns-1s}+db^{2\gamma(pure)}_{ns-1s}+db^{2\gamma(interference)}_{ns-1s}=\frac{1}{\Gamma_{ns}}(dw^{2\gamma(cascade)}_{ns-1s}+dw^{2\gamma(pure)}_{ns-1s}+dw^{2\gamma(interference)}_{ns-1s})\;.
\end{eqnarray}

Combining now the formulas (\ref{54}), (\ref{55}) we arrive at the expression (\ref{4}) presented in the Introduction. The integration of Eq. (\ref{56}) over the remaining frequency will give the total branching ratio
\begin{eqnarray}
\label{57}
b^{2\gamma}_{ns-1s}=\frac{W^{2\gamma}_{ns-1s}}{\Gamma_{ns}}\;.
\end{eqnarray}

Note that this last integration according to Eq. (\ref{54a}) should be done within the interval ($ 0 $, $ \omega_{max} $) since now no pole approximation can be used.

The second term in the curly brackets in Eq. (\ref{53}) is represented by the Feynman graph Fig. 3a (with the change of the photons $ \omega_{f_1}\leftrightarrows\omega_{f_2} $) and corresponds to the resonance 2 in Eq. (\ref{51}), i.e. to the lower link of cascade. To obtain the Lorentz profile for this lower link we have to integrate over the frequency of the first emitted photon, i.e. again over $ \omega_{f_2} $ after taking the square modulus of this term and the factor outside the curly brackets in Eq. (\ref{53}). Replacing notation $ \omega_{f_1} $ to $ \omega  $ we obtain the cascade contribution (resonance 2) to the differential branching ratio:
\begin{eqnarray}
\label{58}
db^{2\gamma(resonance 2)}_{3s-2p-1s}(\omega)=\frac{1}{2\pi}\frac{1}{\Gamma_{3s}}\frac{W_{3s-2p}^{1\gamma}(\omega^{res. 1})W_{2p-1s}^{1\gamma}(\omega^{res. 2})d\omega}{(\omega-\omega^{res. 2})^2+\frac{1}{4}\Gamma_{2p}^2}\;.
\end{eqnarray}
Combining the formulas (\ref{55}) and (\ref{58}) we arrive at the expression (\ref{5}) presented in the Introduction.

The interference between two terms in Eq. (\ref{53}) should not be taken into account since these two terms correspond to the resonances located far from each other: at the distance $ \omega_{max} $ in the frequency scale.
\section{"Pure two-photon" and interference contributions}

Returning to Eq. (\ref{49}) we consider this expression with the state $ 2p $ excluded from the summation over $ n_1 $ as a "pure two-photon" contribution to the transition amplitude $ 3s-1s $. This corresponds to the "pole approximation" employed in Section IV for the description of the cascade contribution: extension of the of the first frequency integration over the interval ($ -\infty,\infty $). In \cite{14} the more general approach was developed, when the resonances were regularized only within the "windows" of the different breadth. Then the $ 2p $ state should be eliminated from the sum over $ n_1 $ only within "windows". The "pole approximation" corresponds to the window breadth $[\omega]=\infty$. This case was considered in \cite{8}.

Since the energy denominators (apart from the factor outside the curly brackets in (\ref{49})) now become nonsingular we can employ the energy conservation law Eq. (\ref{38}) to replace the frequency $ \omega_{f_1} $ in the second denominator in curly brackets in Eq. (\ref{49}) by $ \omega_{f_1}=\omega_{max}-\omega_{f_2} $. Then the "pure two-photon" contribution to the amplitude becomes

\begin{eqnarray}
\label{59}
U^{em., pure}_{3s-1s}=e^2\sum\limits_{n_1\neq 2p}\left\lbrace \frac{(\gamma_{\mu_1}A_{\mu_1}^{*(\vec{k}_{f_2}\vec{e}_{f_2})})_{1sn_1}(\gamma_{\mu_2}A_{\mu_2}^{*(\vec{k}_{f_1}\vec{e}_{f_1})})_{n_13s}}{\omega_{f_2}+E_{1s}-E_{n_1}}+\frac{(\gamma_{\mu_1}A_{\mu_1}^{*(\vec{k}_{f_1}\vec{e}_{f_1})})_{1sn_1}(\gamma_{\mu_2}A_{\mu_2}^{*(\vec{k}_{f_2}\vec{e}_{f_2})})_{n_13s}}{E_{3s}-\omega_{f_2}-E_{n_1}}\right\rbrace\times\\\nonumber \frac{1}{\omega_{f_2}+\omega_{f_1}+E_{1s}-E_{3s}+\frac{i}{2}\Gamma_{3s}}\;.
\end{eqnarray}
Taking Eq. (\ref{59}) by square modulus, integrating over the directions of the emitted photons, summing over the polarizations, integrating over $ \omega_{f_1} $, and changing the notation $ \omega_{f_2}=\omega $ results
\begin{eqnarray}
\label{60}
db^{2\gamma(pure)}_{3s-1s}(\omega)=e^4\omega^2(\omega_0-\omega)^2\sum\limits_{\vec{e}_{f_1}}\sum\limits_{\vec{e}_{f_2}}\int \frac{d\vec{\nu_{f_1}}}{(2\pi)^3}\frac{d\vec{\nu_{f_2}}}{(2\pi)^3}\times\nonumber\\\left|\sum\limits_{n_1\neq 2p}\left\lbrace \frac{(\gamma_{\mu_1}A_{\mu_1}^{*(\vec{k}_{f_2}\vec{e}_{f_2})})_{1sn_1}(\gamma_{\mu_2}A_{\mu_2}^{*(\vec{k}_{f_1}\vec{e}_{f_1})})_{n_13s}}{\omega+E_{1s}-E_{n_1}}+\frac{(\gamma_{\mu_1}A_{\mu_1}^{*(\vec{k}_{f_1}\vec{e}_{f_1})})_{1sn_1}(\gamma_{\mu_2}A_{\mu_2}^{*(\vec{k}_{f_2}\vec{e}_{f_2})})_{n_13s}}{E_{3s}-\omega-E_{n_1}}\right\rbrace\;\right|^2 \frac{1}{\Gamma_{3s}}d\omega\;.
\end{eqnarray}
Then, according to Eq. (\ref{56}) the "pure two-photon" contribution to the differential transition rate is
\begin{eqnarray}
\label{61}
dW^{2\gamma(pure)}_{3s-1s}(\omega)=e^4\omega^2(\omega_0-\omega)^2\sum\limits_{\vec{e}_1}\sum\limits_{\vec{e}_2}\int \frac{d\vec{\nu_{f_1}}}{(2\pi)^3}\frac{d\vec{\nu_{f_2}}}{(2\pi)^3}\times\nonumber\\\left|\sum\limits_{n_1\neq 2p}\left\lbrace \frac{(\gamma_{\mu_1}A_{\mu_1}^{*(\vec{k}_{f_2}\vec{e}_{f_2})})_{1sn_1}(\gamma_{\mu_2}A_{\mu_2}^{*(\vec{k}_{f_1}\vec{e}_{f_1})})_{n_13s}}{\omega+E_{1s}-E_{n_1}}+\frac{(\gamma_{\mu_1}A_{\mu_1}^{*(\vec{k}_{f_1}\vec{e}_{f_1})})_{1sn_1}(\gamma_{\mu_2}A_{\mu_2}^{*(\vec{k}_{f_2}\vec{e}_{f_2})})_{n_13s}}{E_{3s}-\omega-E_{n_1}}\right\rbrace\;\right|^2 d\omega\;.
\end{eqnarray}
Now, using Eq. (\ref{53}) and Eq. (\ref{59}) we can write down the interference contribution to the differential branching ratio as
\begin{eqnarray}
\label{62} 
db^{2\gamma(interference)}_{3s-1s}=2Re\sum\limits_{\vec{e}_1}\sum\limits_{\vec{e}_2}\int \frac{d\vec{\nu_{f_1}}}{(2\pi)^3}\int\frac{d\vec{\nu_{f_2}}}{(2\pi)^3}\int d\omega_{f_1}\omega_{f_1}^2\omega_{f_2}^2U^{em(pure)*}_{3s-1s}U^{em.(cascade)}=\\\nonumber =2Re\;e^4 \sum\limits_{\vec{e}_1}\sum\limits_{\vec{e}_2}\int\frac{d\vec{\nu_{f_1}}}{(2\pi)^3}\int \frac{d\vec{\nu_{f_2}}}{(2\pi)^3}\int d\omega_{f_1}\omega_{f_1}^2\omega_{f_2}^2\times\\\nonumber\left( \sum\limits_{n_1\neq 2p}\left\lbrace\frac{(\gamma_{\mu_1}A_{\mu_1}^{*(\vec{k}_{f_2}\vec{e}_{f_2})})_{1sn_1}(\gamma_{\mu_2}A_{\mu_2}^{*(\vec{k}_{f_1}\vec{e}_{f_1})})_{n_13s}}{\omega_{f_2}+E_{1s}-E_{n_1}}+\frac{(\gamma_{\mu_1}A_{\mu_1}^{*(\vec{k}_{f_1}\vec{e}_{f_1})})_{1sn_1}(\gamma_{\mu_2}A_{\mu_2}^{*(\vec{k}_{f_2}\vec{e}_{f_2})})_{n_13s}}{E_{3s}-E_{n_1}-\omega_{f_2}}\right\rbrace\right)^*\times\\\nonumber\left\lbrace \frac{(\gamma_{\mu_1}A_{\mu_1}^{*(\vec{k}_{f_2}\vec{e}_{f_2})})_{1sn_1}(\gamma_{\mu_2}A_{\mu_2}^{*(\vec{k}_{f_1}\vec{e}_{f_1})})_{n_13s}}{\omega_{f_2}+E_{1s}-E_{2p}+\frac{i}{2}\Gamma_{2p}}+\frac{(\gamma_{\mu_1}A_{\mu_1}^{*(\vec{k}_{f_1}\vec{e}_{f_1})})_{1sn_1}(\gamma_{\mu_2}A_{\mu_2}^{*(\vec{k}_{f_2}\vec{e}_{f_2})})_{n_13s}}{\omega_{f_1}+E_{1s}-E_{2p}+\frac{i}{2}\Gamma_{2p}}\right\rbrace\times\\\nonumber\frac{d\omega_{f_2}}{(\omega_{f_1}+\omega_{f_2}+E_{1s}-E_{3s})^2+\frac{1}{4}\Gamma_{3s}^2}\;.
\end{eqnarray}

The integration in the complex $ \omega_{f_1} $ plane can be extended over the entire interval $ -\infty\leqslant\omega_{f_1}\leqslant+\infty $ since only the pole term contributes; then we have to take the real part of the expression obtained.

The "pure two-photon" amplitude in Eq. (\ref{62}) we can assume to be pure real.

Then using Eq. (\ref{55}) for the interference contribution to the differential two-photon transition rate $ 3s-1s $ we find (changing notation $ \omega_{f_2} $ to $ \omega $)

\begin{eqnarray}
\label{63}
dW^{2\gamma(interference)}_{3s-1s}(\omega)=\frac{2(\omega-\omega^{res2})F_1^{3s}(\omega)}{(\omega-\omega^{res2})^2+\frac{1}{4}(\Gamma_{3s}+\Gamma_{2p})^2}d\omega+\frac{2(\omega-\omega^{res1})F_2^{3s}(\omega)}{(\omega-\omega^{res1})^2+\frac{1}{4}\Gamma_{2p}^2}d\omega\;.
\end{eqnarray}

\begin{eqnarray}
\label{64}
F^{3s}_i(\omega)=\sum\limits_{\vec{e}_{f_1}}\sum\limits_{\vec{e}_{f_2}}\int \frac{d\vec{\nu_{f_1}}}{(2\pi)^3}\int\frac{d\vec{\nu_{f_2}}}{(2\pi)^3}f^{3s}(\omega)\varphi_{i}^{3s}, i=1,2\;.
\end{eqnarray}

\begin{eqnarray}
\label{65}
f^{3s}=\omega^2(\omega^{3s}_0-\omega)^2\sum\limits_{n_1\neq 2p}\left\lbrace \frac{(\gamma_{\mu_1}A_{\mu_1}^{*(\vec{k}_{f_2}\vec{e}_{f_2})})_{1sn_1}(\gamma_{\mu_2}A_{\mu_2}^{*(\vec{k}_{f_1}\vec{e}_{f_1})})_{n_13s}}{(\omega+E_{1s}-E_{n_1})}+\frac{(\gamma_{\mu_1}A_{\mu_1}^{*(\vec{k}_{f_1}\vec{e}_{f_1})})_{1sn_1}(\gamma_{\mu_2}A_{\mu_2}^{*(\vec{k}_{f_2}\vec{e}_{f_2})})_{n_13s}}{(E_{3s}-E_{n_1}-\omega)}\right\rbrace\;,
\end{eqnarray}
\begin{eqnarray}
\label{66}
\varphi_{1}^{3s}=(\gamma_{\mu_1}A_{\mu_1}^{*(\vec{k}_{f_2}\vec{e}_{f_2})})_{1s2p}(\gamma_{\mu_2}A_{\mu_2}^{*(\vec{k}_{f_1}\vec{e}_{f_1})})_{2p3s}\;,
\end{eqnarray}

\begin{eqnarray}
\label{67}
\varphi_{2}^{3s}=(\gamma_{\mu_1}A_{\mu_1}^{*(\vec{k}_{f_1}\vec{e}_{f_1})})_{1s2p}(\gamma_{\mu_2}A_{\mu_2}^{*(\vec{k}_{f_2}\vec{e}_{f_2})})_{2p3s}\;.
\end{eqnarray}
In Eqs. (\ref{65}) - (\ref{67}) it is assumed that $ |k_{f_2}|\equiv\omega $, $ |k_{f_1}|\equiv(\omega^{3s}_0-\omega) $. Unlike the cascade contribution, the dependence on $ \Gamma_{3s}+\Gamma_{2p} $ in Eq. (\ref{63}) is essential and does not disappear after the insertion of Eq. (\ref{63}) in Eq. (\ref{1}). The employment of the correct expression Eq. (\ref{63}) for the interference contribution is important for the evaluating of the CMB properties in astrophysics.

\section{Two-photon $4s-1s$ transition}
Repeating the derivations for the $ 3s-1s $ transition for the case of $ 4s-1s $ transition we present first the contributions for the two cascades: $ 4s-2p-1s $ and $ 4s-3p-1s $:

1) Contribution from the upper link $ 4s-2p $ of the cascade $ 4s-2p-1s $:
\begin{eqnarray}
\label{68}
dW^{2\gamma(resonance 1)}_{4s-2p-1s}=\frac{1}{2\pi}\frac{\Gamma_{4s}+\Gamma_{2p}}{\Gamma_{2p}}\frac{W_{4s-2p}^{1\gamma}(\omega^{res. 1})W_{2p-1s}^{1\gamma}(\omega^{res. 2})d\omega}{(\omega-\omega^{res. 1})^2+\frac{1}{4}(\Gamma_{4s}+\Gamma_{2p})^2}\;,
\end{eqnarray}
where
\begin{eqnarray}
\label{69}
\omega^{res. 1}=E_{4s}-E_{2p},\;\; \omega^{res. 2}=E_{2p}-E_{1s}\;.
\end{eqnarray}
2) Contribution of the lower link $ 2p-1s $ of the cascade $ 4s-2p-1s $
\begin{eqnarray}
\label{70}
dW^{2\gamma(resonance 1)}_{4s-2p-1s}=\frac{W_{4s-2p}^{1\gamma}(\omega^{res. 1})W_{2p-1s}^{1\gamma}(\omega^{res. 2})d\omega}{(\omega-\omega^{res. 1})^2+\frac{1}{4}\Gamma_{2p}^2}\;.
\end{eqnarray}
3) Contribution from the upper link $ 4s-3p $ of the cascade $ 4s-3p-1s $:
\begin{eqnarray}
\label{71}
dW^{2\gamma(resonance 1)}_{4s-3p-1s}=\frac{1}{2\pi}\frac{\Gamma_{4s}+\Gamma_{3p}}{\Gamma_{3p}}\frac{W_{4s-3p}^{1\gamma}(\omega^{res. 3})W_{3p-1s}^{1\gamma}(\omega^{res. 4})d\omega}{(\omega-\omega^{res. 3})^2+\frac{1}{4}(\Gamma_{4s}+\Gamma_{3p})^2}\;,
\end{eqnarray}
where
\begin{eqnarray}
\label{72}
\omega^{res. 3}=E_{4s}-E_{3p},\;\; \omega^{res. 4}=E_{3p}-E_{1s}\;.
\end{eqnarray}
4) Contribution of the lower link $ 3p-1s $ of the cascade $ 4s-3p-1s $
\begin{eqnarray}
\label{73}
dW^{2\gamma(resonance 1)}_{4s-3p-1s}=\frac{W_{4s-3p}^{1\gamma}(\omega^{res. 3})W_{3p-1s}^{1\gamma}(\omega^{res. 4})d\omega}{(\omega-\omega^{res. 4})^2+\frac{1}{4}\Gamma_{3p}^2}\;.
\end{eqnarray}
Insertion of Eqs. (\ref{68}), (\ref{70}), (\ref{71}), (\ref{73}) in Eq. (\ref{1}) yields:
\begin{eqnarray}
\label{74}
dW^{2\gamma(cascade)}_{4s-1s}=\frac{1}{2}\int\limits_0^{\omega_0}\sum\limits_{i=1}^4dW^{2\gamma(resonance\; i)}=W^{1\gamma}_{4s-2p}+\frac{W^{1\gamma}_{3p-1s}}{\Gamma_{3p}}W^{1\gamma}_{4s-3p}=W^{1\gamma}_{4s-2p}+b^{1\gamma}_{3p-1s}W^{1\gamma}_{4s-3p}\;,
\end{eqnarray}
where $\;\omega_0=E_{4s}-E_{1s} $, $ b^{1\gamma}_{3p-1s} $ is the branching ratio for the transition $ 3p-1s $. We took into account that $ \Gamma_{3p}=W^{1\gamma}_{3p-1s}+W^{1\gamma}_{3p-2s} $ and $ b^{1\gamma}_{3p-1s}=\frac{W^{1\gamma}_{3p-1s}}{W^{1\gamma}_{3p-1s}+W^{1\gamma}_{3p-2s}} $.

Hence 
\begin{eqnarray}
\label{75}
W^{2\gamma(cascade)}_{4s-1s}\neq\Gamma_{4s}\;,
\end{eqnarray}
where
\begin{eqnarray}
\label{76}
\Gamma_{4s} =W^{1\gamma}_{4s-2p}+W^{1\gamma}_{4s-3p}\;,
\end{eqnarray}
unlike Eq. (\ref{7a}) in case of $ 3s-1s $ transition.
The "pure two-photon" contribution to the $ 4s-1s $ two-photon differential decay rate looks similar to the $ 3s-1s $ case (see Eq. (\ref{61})):
\begin{eqnarray}
\label{77}
dW^{2\gamma(pure)}_{4s-1s}(\omega)=e^4\omega^2(\omega_0-\omega)^2\sum\limits_{\vec{e}_{f_1}}\sum\limits_{\vec{e}_{f_2}}\int \frac{d\vec{\nu}_{f_1}}{(2\pi)^3}\frac{d\vec{\nu}_{f_2}}{(2\pi)^3}\times\\\nonumber\left|\sum\limits_{n_1\neq 2p,3p}\left\lbrace \frac{(\gamma_{\mu_1}A_{\mu_1}^{*(\vec{k}_{f_2}\vec{e}_{f_2})})_{1sn_1}(\gamma_{\mu_2}A_{\mu_2}^{*(\vec{k_{f_1}}\vec{e_{f_1}})})_{n_14s}}{\omega+E_{1s}-E_{n_1}}+\frac{(\gamma_{\mu_1}A_{\mu_1}^{*(\vec{k}_{f_1}\vec{e}_{f_1})})_{1sn_1}(\gamma_{\mu_2}A_{\mu_2}^{*(\vec{k}_{f_2}\vec{e}_{f_2})})_{n_14s}}{E_{4s}-E_{n_1}-\omega}\right\rbrace\right|^2d\omega\;.
\end{eqnarray}
Interference contribution for the $ 4s-1s $ transition consists of 4 terms, corresponding to the four resonances, presented by Eqs. (\ref{69}), (\ref{72})
\begin{eqnarray}
\label{78}
dW^{2\gamma(interference)}_{4s-1s}(\omega)=\frac{2(\omega-\omega^{res1})F_1^{4s}(\omega)}{(\omega-\omega^{res1})^2+\frac{1}{4}(\Gamma_{4s}+\Gamma_{2p})^2}d\omega+\frac{2(\omega-\omega^{res2})F_2^{4s}(\omega)}{(\omega-\omega^{res2})^2+\frac{1}{4}\Gamma_{2p}^2}d\omega+\nonumber\\+\frac{2(\omega-\omega^{res3})F_3^{4s}(\omega)}{(\omega-\omega^{res3})^2+\frac{1}{4}(\Gamma_{4s}+\Gamma_{3p})^2}d\omega+\frac{2(\omega-\omega^{res4})F_4^{4s}(\omega)}{(\omega-\omega^{res4})^2+\frac{1}{4}\Gamma_{3p}^2}d\omega\;,
\end{eqnarray}
where 
\begin{eqnarray}
\label{79}
F^{4s}_i(\omega)=\sum\limits_{\vec{e}_{f_1}}\sum\limits_{\vec{e}_{f_2}}\int \frac{d\vec{\nu}_{f_1}}{(2\pi)^3}\frac{d\vec{\nu}_{f_2}}{(2\pi)^3}f^{4s}(\omega)\varphi_{i}^{4s},\; i=1,2,3,4\;,
\end{eqnarray}

\begin{eqnarray}
\label{80}
f^{4s}(\omega)=\omega^2(\omega^{4s}_0-\omega)^2\sum\limits_{n_1\neq 2p,3p}\left\lbrace \frac{(\gamma_{\mu_1}A_{\mu_1}^{*(\vec{k}_{f_2}\vec{e}_{f_2})})_{1sn_1}(\gamma_{\mu_2}A_{\mu_2}^{*(\vec{k}_{f_1}\vec{e}_{f_1})})_{n_14s}}{\omega+E_{1s}-E_{n_1}}+\right.\\\nonumber\left.+\frac{(\gamma_{\mu_1}A_{\mu_1}^{*(\vec{k_{f_1}}\vec{e_{f_1}})})_{1sn_1}(\gamma_{\mu_2}A_{\mu_2}^{*(\vec{k}_{f_2}\vec{e}_{f_2})})_{n_14s}}{E_{4s}-E_{n_1}-\omega}\right\rbrace\;,
\end{eqnarray}

\begin{eqnarray}
\label{81}
\varphi_{1}^{4s}=(\gamma_{\mu_1}A_{\mu_1}^{*(\vec{k}_{f_2}\vec{e}_{f_2})})_{1s2p}(\gamma_{\mu_2}A_{\mu_2}^{*(\vec{k}_{f_1}\vec{e}_{f_1})})_{2p4s}\;,
\end{eqnarray}

\begin{eqnarray}
\label{82}
\varphi_{2}^{4s}=(\gamma_{\mu_1}A_{\mu_1}^{*(\vec{k}_{f_1}\vec{e}_{f_1})})_{1s2p}(\gamma_{\mu_2}A_{\mu_2}^{*(\vec{k}_{f_2}\vec{e}_{f_2})})_{2p4s}\;,
\end{eqnarray}

\begin{eqnarray}
\label{81}
\varphi_{3}^{4s}=(\gamma_{\mu_1}A_{\mu_1}^{*(\vec{k}_{f_2}\vec{e}_{f_2})})_{1s3p}(\gamma_{\mu_2}A_{\mu_2}^{*(\vec{k}_{f_1}\vec{e}_{f_1})})_{3p4s}\;,
\end{eqnarray}

\begin{eqnarray}
\label{82}
\varphi_{4}^{4s}=(\gamma_{\mu_1}A_{\mu_1}^{*(\vec{k}_{f_1}\vec{e}_{f_1})})_{1s3p}(\gamma_{\mu_2}A_{\mu_2}^{*(\vec{k}_{f_2}\vec{e}_{f_2})})_{3p4s}\;.
\end{eqnarray}

\section{Three-photon $3p\rightarrow 1s+3\gamma $ transition}

In this section we consider the 3-photon transitions, taking as an example $3p\rightarrow 1s+3\gamma $ transition. The main decay channels of the $3p$ level are $ 3p\rightarrow 1s+\gamma $ and $ 3p\rightarrow 2s+\gamma $. Therefore as a resonance scattering process in this case we can choose the one-photon absorption and three-photon emission process depicted in Fig. 4. For the transition $3p\rightarrow 1s+3\gamma $ there are two cascades, containing two-photon links: $ 3p\rightarrow 2p+2\gamma\rightarrow 1s+3\gamma $ and $3p\rightarrow 2s+\gamma\rightarrow 1s+3\gamma$.

In case of the $ 3 $-photon transition we will take into account only cascade contribution. This cascade contribution contains necessarily one "pure two-photon" link ($ 3p-2p $ or $ 2s-1s $) and is, therefore of the same order of magnitude as the "pure two-photon" contribution to the $ 3s-1s $ transition. The  "pure 3-photon" transitions and the corresponding interference terms are essentially smaller than cascade contributions and can be neglected in the  "two-photon" approximation \cite{16}. In this sense the situation differs from the situation in $ 3s-1s $ two-photon decay when we were interested in the "pure two-photon" and interference terms.

The derivation similar to the two-photon case gives the following expression for the 3-photon emission amplitude $ 3p-1s $ in the resonance approximation:
\begin{eqnarray}
\label{83}
U^{em.3\gamma}_{3p-1s}=e^3\sum\limits_{n_1n_2}\frac{(\gamma_{\mu_1}A^{*\vec{k}_{f_3}\vec{e}_{f_3}}_{\mu_1})_{1sn_1}(\gamma_{\mu_2}A^{*\vec{k}_{f_2}\vec{e}_{f_2}}_{\mu_2})_{n_1n_2}(\gamma_{\mu_3}A^{*\vec{k}_{f_1}\vec{e}_{f_1}}_{\mu_3})_{n_23p}}{(E_{1s}-E_{n_1}+\omega_{f_3})(E_{1s}-E_{n_2}+\omega_{f_3}+\omega_{f_2})}\\\nonumber\times\frac{1}{E_{1s}-E_{3p}+\omega_{f_3}+\omega_{f_2}+\omega_{f_1}+\frac{i}{2}\Gamma_{3p}}\;.
\end{eqnarray}
In Eq. (\ref{83}) we summed already all the self-energy insertions in the lower electron propagator as is shown in Fig. 4. An exact energy conservation law in case of the process Fig. 4 is
\begin{eqnarray}
\label{83a}
\omega_i=\omega_{f_1}+\omega_{f_2}+\omega_{f_3}\;.
\end{eqnarray}
The resonance condition and the approximate energy conservation law in case of 3-photon decay looks like 
\begin{eqnarray}
\label{84}
|\omega_i-(E_{3p}-E_{1s})|=|\omega_{f_1}+\omega_{f_2}+\omega_{f_3}-(E_{3p}-E_{1s})|\leqslant \Gamma_{3p}\;.
\end{eqnarray}
To fix the cascade $3p\rightarrow 2s+\gamma \rightarrow1s+3\gamma  $ contribution  we set $ n_2=2s $ in Eq. (\ref{83}). This results 
\begin{eqnarray}
\label{85}
U^{em.3\gamma}_{3p-1s}=e^3\frac{1}{E_{1s}-E_{3p}+\omega_{f_3}+\omega_{f_2}+\omega_{f_1}+\frac{i}{2}\Gamma_{3p}}\frac{(\gamma_{\mu_1}A^{*\vec{k}_{f_1}\vec{e}_{f_1}}_{\mu_1})_{3p2s}}{(E_{1s}-E_{2s}+\omega_{f_3}+\omega_{f_2}+\frac{i}{2}\Gamma_{2s})}\times\\\nonumber\sum\limits_{n_1}\frac{(\gamma_{\mu_2}A^{*\vec{k}_{f_2}\vec{e}_{f_2}}_{\mu_2})_{2sn_1}(\gamma_{\mu_2}A^{*\vec{k}_{f_3}\vec{e}_{f_3}}_{\mu_3})_{n_11s}}{(E_{1s}-E_{n_1}+\omega_{f_3})}+(perm)\;.
\end{eqnarray}
In Eq. (\ref{85}) we should include the contribution of the Feynman graphs with all the permutations of the photon lines (perm).

Now we have to take the right-hand side of Eq. (\ref{85}) by square modulus, to integrate over the emitted photon directions and to sum over the photon polarizations. Then we have to integrate over the photon frequencies $ \omega_{f_1} $, $ \omega_{f_2} $, $ \omega_{f_3} $ taking into account the condition Eq. (\ref{84}). However when we integrate the contribution of the cascade $ 3p-2s-1s $ we have to take into account that the frequency $ \omega_{f_1} $ is fixed by the resonance condition
\begin{equation}
\label{86}
| \omega_{f_1} -(E_{3p}-E_{2s})|\leqslant \Gamma_{3p}+\Gamma_{2s}\;.
\end{equation}
Inserting Eq. (\ref{86}) in the approximate conservation law (\ref{84}) we obtain the approximate equality
\begin{eqnarray}
\label{87}
\omega_{f_2}+\omega_{f_3}=E_{2s}-E_{1s}\;.
\end{eqnarray}
The integration over $ \omega_{f_2}$, $ \omega_{f_3} $ should be performed in the following way
\begin{eqnarray}
\label{88}
\int\limits_0^{E_{2s}-E_{1s}}d\omega_{f_2}\int\limits_0^{\omega_{f_2}}d\omega_{f_3}=\frac{1}{2}\int\limits_0^{E_{2s}-E_{1s}}d\omega_{f_2}\int\limits_0^{E_{2s}-E_{1s}}d\omega_{f_3}\;.
\end{eqnarray}
Eq. (\ref{88}) holds after symmetrization of Eq. (\ref{86}) via the permutation of the photons with the frequencies $ \omega_{f_2} $, $ \omega_{f_3} $.
The integration over two frequencies (e.g. over $ \omega_{f_1} $ and $ \omega_{f_2} $ in Eq. (\ref{85})) can be always extended to the interval  [$-\infty$, $+\infty$] since the pole approximation can be used. The third integration over $ \omega_{f_3} $ in Eq. (\ref{85}) according to Eq. (\ref{88}) should be performed over the finite interval [0, $E_{2s}-E_{1s}$]. The integration yields  
\begin{eqnarray}
\label{89}
b^{3\gamma}_{3p-1s}(3p-2s-1s)=2\pi e^3 \frac{\Gamma_{3p}+\Gamma_{2s}}{\Gamma_{3p}\Gamma_{2s}}\int\limits_{-\infty}^{\infty}\omega^2_{_{f_1}}d\omega_{f_1}\sum\limits_{\vec{e}_{f_1}}\int \frac{d\vec{\nu}_{f_1}}{(2\pi)^3}\frac{|U^{em.1\gamma}(3p-2s-1s)|}{(E_{3p}-E_{2s}-\omega_{f_1})^2+\frac{1}{4}(\Gamma_{3p}+\Gamma_{2s})^2}\times\\\nonumber\frac{1}{2}\int\limits_0^{\omega_{max}}\omega^2_{f_3}(\omega_{max}-\omega_{f_3})^2d\omega_{f_3}\sum\limits_{\vec{e}_{f_2}}\sum\limits_{\vec{e}_{f_3}}\frac{d\vec{\nu}_{f_2}}{(2\pi)^3}\frac{d\vec{\nu}_{f_3}}{(2\pi)^3}\sum\limits_{n_1}\frac{(\gamma_{\mu_2}A_{\mu_2}^{*(\vec{k}_{f_3}\vec{e}_{f_3})})_{1sn_1}(\gamma_{\mu_3}A_{\mu_3}^{*(\vec{k}_{f_2}\vec{e}_{f_2})})_{n_12s}}{E_{1s}-E_{n_1}+\omega_{f_3}}=\frac{W^{1\gamma}_{3p-2s}W^{2\gamma}_{2s-1s}}{\Gamma_{3p}\Gamma_{2s}}\;,
\end{eqnarray}
where $ \omega_{max}=E_{2s}-E_{1s} $.

The physical sense of the dimensionless quantity $ b^{3\gamma}_{3p-1s}(3p-2s-1s) $ should be discussed specially. This quantity should define the $ 3\gamma $ transition rate $ 3p-1s $
 via the channel $ 3p\rightarrow2s+\gamma \rightarrow 1s+3\gamma $. This transition rate is very small compared to the main decay channel for the $ 3p $ state, i.e. $ W^{1\gamma}_{3p-1s} $. We assume that the quantity $ b^{3\gamma}_{3p-1s}(3p-2s-1s) $ is the branching ratio for the $ 3 $-photon transition rate $ W^{3\gamma}_{3p-1s}(3p-2s-1s) $ to the direct two-photon transition rate $ W^{2\gamma}=\Gamma_{2s} $. Then from Eq. (\ref{89}) it follows
\begin{eqnarray}
\label{90}
W^{3\gamma}_{3p-1s}(3p-2s-1s)=\frac{W^{1\gamma}_{3p-2s}}{\Gamma_{3p}}W^{2\gamma}_{2s-1s}\;.
\end{eqnarray}
In the same way the contribution of the $ 3 $-photon cascade $ 3p-2p+2\gamma\rightarrow 1s+3\gamma $ can be analysed. The final result looks like 
\begin{eqnarray}
\label{91}
b^{3\gamma}(3p-2p-1s)=\frac{W^{1\gamma}_{2p-1s}}{\Gamma_{2p}\Gamma_{3p}}W^{2\gamma}_{3p-2p}\;.
\end{eqnarray}
Unlike $b^{3\gamma}(3p-2s-1s)$ the quantity Eq. (\ref{91}) should be considered as the branching ratio of the transition rate via channel $ 3p-2p-1s $ to the total width of the $ 3p $ level, i.e. $ \Gamma_{3p} $. Then
\begin{eqnarray}
\label{92}
W^{3\gamma}_{3p-1s}(3p-2p-1s)=\frac{W^{1\gamma}_{2p-1s}}{\Gamma_{2p}}W^{2\gamma}_{3p-2p}\;.
\end{eqnarray}
In the equation in \cite{16}, corresponding to the Eq. (\ref{92}), the factor $ W^{1\gamma}_{2p-1s}/\Gamma_{2p}=1$ was omitted. Here we keep it to demonstrate that the transition channel $ 3p-2p-1s $ is a $ 3 $-photon channel. Total expression for the transition rate $ 3p-1s $ ($ 1 $-photon and $ 3 $-photon) is
\begin{eqnarray}
\label{93}
W^{(3\gamma)}_{3p-1s}=W^{(1\gamma)}_{3p,1s}+\frac{W^{(1\gamma)}_{2p,1s}}{\Gamma_{2p}}W^{(2\gamma)}_{3p,2p}+\frac{W^{(1\gamma)}_{3p,2s}}{\Gamma_{3p}}W^{(2\gamma)}_{2s,1s}.
\end{eqnarray}
This expression coincides with one derived in \cite{16} up to the coefficients before the second and third terms in the right-hand side of Eq. (\ref{94}). In \cite{16} this coefficients were evaluated incorrectly and were equal $ 3/4 $. This error was notices by the authors of \cite{15}. However an expression for $ W_{3p-1s} $ given in \cite{15} is different form Eq. (\ref{93}). The equation in \cite{15} reads
\begin{eqnarray}
\label{94}
W_{3p-1s}=\Gamma^{1\gamma}_{3p-1s}+\Gamma^{2\gamma}_{3p-2p}+\Gamma^{1\gamma}_{3p-2s}\;,
\end{eqnarray}
where $ \Gamma^{1\gamma}_{3p-1s} $, $ \Gamma^{1\gamma}_{3p-2s} $ and $ \Gamma^{2\gamma}_{3p-2p} $  are the partial widths for different decay channels. Since $ \Gamma^{1\gamma}_{3p-1s}=W^{1\gamma}_{3p-1s} $, $ \Gamma^{2\gamma}_{3p-2p}=W^{2\gamma}_{3p-2p} $, $ \Gamma^{1\gamma}_{3p-2s}=W^{1\gamma}_{3p-2s} $, Eq. (\ref{94}) differs form Eq. (\ref{93}) by the last term in the right hand side. Eq. (\ref{94}) represents the total width of the $ 3p $ level. As in cases $ 3s-1s $, $ 4s-1s $ decays $  W^{1\gamma}_{3p-1s}$ differs from the $ \Gamma_{3p} $ ($ W_{3p-1s}\neq\Gamma_{3p} $). In case of $ 3s $ level relative difference between $ W_{3s-1s} $ and $ \Gamma_{3s} $   was quite small: about $ 10^{-6} $. This difference became of order of 1 for $ 4s-1s $. For $ 3p-1s $ decay it is again of the order of 1, though the values of the last term in Eqs. (\ref{93}) and (\ref{94}) differ by many orders of magnitude. Within our treatment Eq. (\ref{94}) arises when we consider Eq. (\ref{89}) as a branching ratio of the decay channel $ 3p\rightarrow 2s+\gamma $ to the total width $ \Gamma_{3p} $. Then, multiplying Eq. (\ref{89}) by $ \Gamma_{3p} $ and setting $ W^{2\gamma}_{2s-1s}/\Gamma_{2s}=1 $ we arrive at 
\begin{eqnarray}
\label{94a}
\Gamma_{3p}=\Gamma^{1\gamma}_{3p-1s}+\Gamma^{2\gamma}_{3p-2s}\;.
\end{eqnarray}

\section{Conclusions}
In this paper we analyzed the problem of the multiphoton transitions with cascades taking as an example the two-photon $  3s\rightarrow 1s+2\gamma$, $ 4s\rightarrow 1s+2\gamma $ transitions and the three-photon $ 3p\rightarrow 1s+3\gamma $ transition. We proved that the regularization of the singularities in the expressions for the cascade contributions to the transition rates includes the widths of the  both initial and intermediate states. This may be important for the the astrophysical studies of the cosmological recombination and, consequently to the connection between the process of the radiation escape from the matter in the recombination epoch and the recent studies of the properties of CMB.

It should be stressed that according to our Eq. (\ref{35}), in principle, the absolute transition probability from any excited state to the ground state cannot be evaluated with the help of the nondiagonal $ S $-matrix elements, since the transition probability turns to zero. However it does not mean that the standard procedure of the evaluation of transition rates via the nondiagonal $ S $-matrix elements is inapplicable. The standard procedure works in all cases when it is not necessary to take into account the total width of the decaying state. When the total width of the decaying state becomes important we need to employ the Low procedure as described in the present paper.

The necessity of this procedure arises when we need to describe the Lorentz profiles for the decay processes of the excited states or when we want to regularize properly the expressions for the multiphoton transition probabilities with cascades. In principle, it should be possible to derive the correct results within the phenomenological QM approach as well. However, for this purpose one should follow the procedure described, for example, in \cite{27} for the one-photon decay of the level, possessing  the width: the solution of the non-stationary Schr\"{o}dinger equation with appropriate initial conditions. This derivation should be extended to the case of the two-photon decay.

\section*{Acknowledgements}
The authors wish to thank Dr. O. Yu. Andreev for many helpful discussions. This work was supported by RFBR grant 11-02-00168a. T. Z. acknowledges support by the non-profit Foundation "Dynasty" (Moscow). 

\setcounter{equation}{0}
\renewcommand{\theequation}%
{A.\arabic{equation}}

\section*{Appendix A: Derivation of Eqs. (\ref{54}), (\ref{58})}
Taking the first term in the curly brackets in Eq. (\ref{53}) together with the factor outside the brackets by square modulus, integrating over the emitted photons directions and introducing the shorthand notations
\begin{eqnarray}
E_{2p}-E_{1s}=\omega^{res. 2}\equiv \Delta E_A\;,
\end{eqnarray}
\begin{eqnarray}
E_{3s}-E_{1s}\equiv\omega_0\equiv \Delta E_B\;,
\end{eqnarray}
we define the double differential branching ratio as 
\begin{eqnarray}
db^{2\gamma(resonance\;1)}_{3s-2p-1s}=\frac{1}{(2\pi)^2}\frac{W^{1\gamma}_{3s-2p}(\omega^{res.1})W^{1\gamma}_{2p-1s}(\omega^{res.2})}{[\Delta E_A-\omega_{f_2}-\frac{i}{2}\Gamma_{2p}][\Delta E_A-\omega_{f_2}+\frac{i}{2}\Gamma_{2p}]}\times\\\nonumber\frac{d\omega_{f_1}d\omega_{f_2}}{[\Delta E_B-\omega_{f_1}-\omega_{f_2}-\frac{i}{2}\Gamma_{3s}][\Delta E_B-\omega_{f_1}-\omega_{f_2}+\frac{i}{2}\Gamma_{3s}]}\;.
\end{eqnarray}
Using Cauchy theorem we integrate over $ \omega_{f_2} $ in the lower half-plane where the poles are:
\begin{eqnarray}
\omega^{(1)}_{f_2}=\Delta E_A-\frac{i}{2}\Gamma_{2p}\;,
\end{eqnarray}
\begin{eqnarray}
\omega^{(2)}_{f_2}=\Delta E_B-\omega_{f_1}-\frac{i}{2}\Gamma_{3s}\;.
\end{eqnarray}
The integration results:

\begin{eqnarray}
db^{2\gamma(resonance\;1)}_{3s-2p-1s}=\frac{1}{2\pi}W^{1\gamma}_{3s-2p}(\omega^{res.1})W^{1\gamma}_{2p-1s}(\omega^{res.2})\times\\\nonumber\left(\frac{1}{\Gamma_{2p}[\Delta E_B-\Delta E_A-\omega_{f_1}+\frac{i}{2}(\Gamma_{2p}-\Gamma_{3s})][\Delta E_B-\Delta E_A-\omega_{f_1}+\frac{i}{2}(\Gamma_{2p}+\Gamma_{3s})]}\right.+\\\nonumber\left.+\frac{1}{\Gamma_{3s}[\Delta E_A-\Delta E_B+\omega_{f_1}-\frac{i}{2}(\Gamma_{2p}-\Gamma_{3s})][\Delta E_A-\Delta E_B+\omega_{f_1}+\frac{i}{2}(\Gamma_{2p}+\Gamma_{3s})]}\right)  d\omega_1\;.
\end{eqnarray}
Algebraic transformations then lead to
\begin{eqnarray}
db^{2\gamma(resonance\;1)}_{3s-2p-1s}=\frac{1}{2\pi}\frac{W^{1\gamma}_{3s-2p}(\omega^{res.1})W^{1\gamma}_{2p-1s}(\omega^{res.2})}{\Gamma_{2p}\Gamma_{3s}}\times\\\nonumber\frac{\Gamma_{3s}[\Delta E_B-\Delta E_A-\omega_{f_1}-\frac{i}{2}(\Gamma_{2p}+\Gamma_{3s})]+\Gamma_{2p}[\Delta E_B-\Delta E_A-\omega_{f_1}+\frac{i}{2}(\Gamma_{2p}+\Gamma_{3s})]}{[ \Delta E_B-\Delta E_A-\omega_{f_1}+\frac{i}{2}(\Gamma_{2p}-\Gamma_{3s})][(\Delta E_B-\Delta E_A-\omega_{f_1})^2+\frac{1}{4}(\Gamma_{2p}+\Gamma_{3s})^2]}=\\\nonumber=\frac{1}{2\pi}\frac{W^{1\gamma}_{3s-2p}(\omega^{res.1})W^{1\gamma}_{2p-1s}(\omega^{res.2})}{\Gamma_{2p}\Gamma_{3s}}\times\\\nonumber\frac{(\Gamma_{3s}+\Gamma_{2p})[\Delta E_B-\Delta E_A-\omega_{f_1}+\frac{i}{2}(\Gamma_{2p}-\Gamma_{3s})]}{[\Delta E_B-\Delta E_A-\omega_{f_1}+\frac{i}{2}(\Gamma_{2p}-\Gamma_{3s})][(\Delta E_B-\Delta E_A-\omega_{f_1})^2+\frac{1}{4}(\Gamma_{2p}+\Gamma_{3s})^2]}\;.
\end{eqnarray}
After the cancellation of the factor $ [\Delta E_B-\Delta E_A-\omega_{f_1}+\frac{i}{2}(\Gamma_{2p}-\Gamma_{3s})] $ in the numerator and the denominator of Eq. (A.7) we arrive of the expression (\ref{54}) given in the text.

To obtain Eq. (\ref{58}) we have to take the second term in the curly brackets in Eq. (\ref{53}), to integrate it over the directions of the emitted photons and to sum over the emitted photons polarizations. This gives
\begin{eqnarray}
db^{2\gamma(resonance\;2)}_{3s-2p-1s}=\frac{1}{2\pi}\frac{W^{1\gamma}_{3s-2p}(\omega^{res.1})W^{1\gamma}_{2p-1s}(\omega^{res.2})d\omega_{f_1}d\omega_{f_2}}{[(\Delta E_A-\omega_{f_1})^2+\frac{1}{4}\Gamma^2_{2p}][(\Delta E_B-\omega_{f_1}-\omega_{f_2})^2+\frac{1}{4}\Gamma^2_{3s}]}\;.
\end{eqnarray}
We again integrate over $ \omega_{f_2} $ in the complex plane obtaining immediately the result Eq. (\ref{58}).

\newpage
\begin{figure}[h!]
  \centering
\includegraphics[scale=0.4]{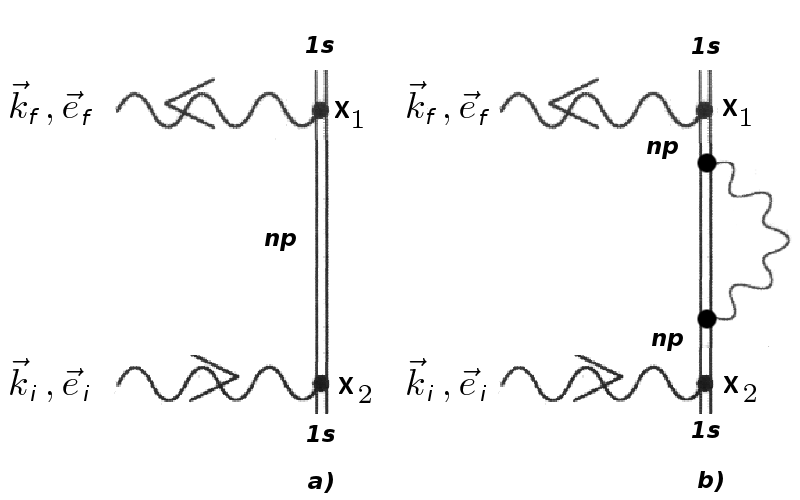}
  \caption{Feynman graph describing the resonant photon scattering on the ground state of hydrogen atom. In FIG. 1$ a $ the basic process of the resonant scattering with the excitation of $ np $ state is depicted. In FIG. 1$ b $ the electron self-energy insertion in the propagator is made. The double solid lines denote the electron in the field of the nucleus (Furry picture of QED), the wavy lines denote the absorbed, emitted and virtual photons.}
\end{figure}
\begin{figure}[h!]
  \centering
\includegraphics[scale=0.4]{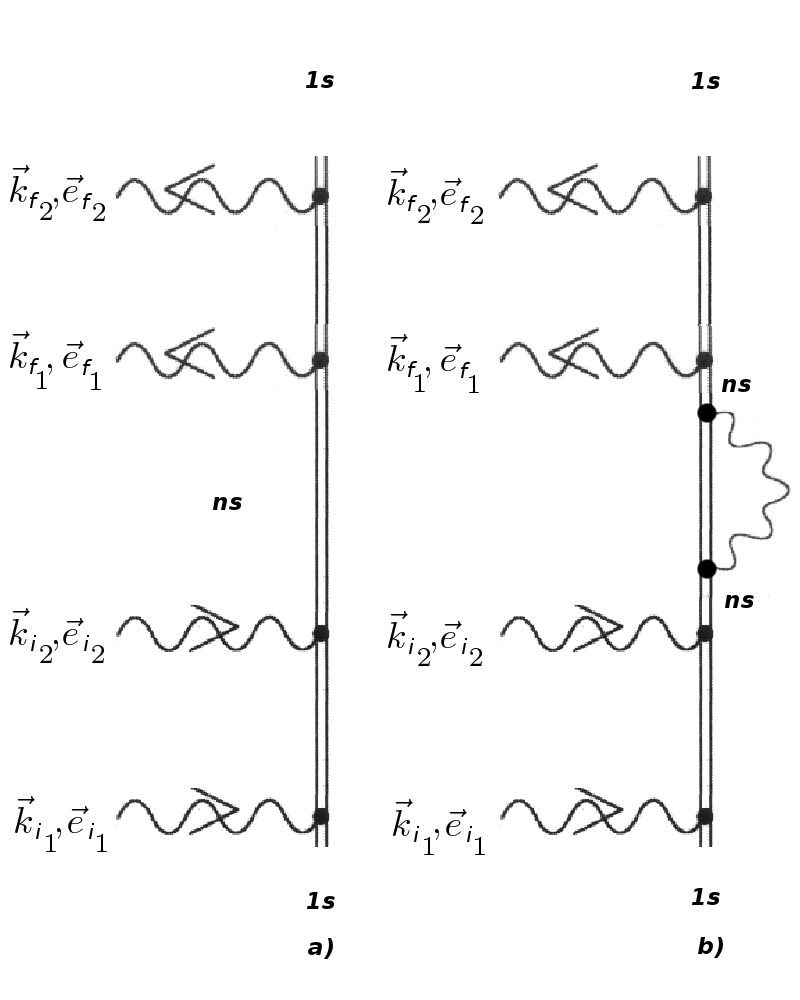}
  \caption{Feynman graph describing the two-photon resonant scattering on the ground state of hydrogen atom, with the excitation of $ ns $ $ (n>2) $ state and the resonance condition $ \omega_1+\omega_2=E_{ns}-E_{1s} $. In FIG. 2$ a $ the basic process of the resonant scattering with the excitation of the $ ns $ state is depicted. In FIG. 2$ b $ the electron self-energy insertion in the central electron propagator is made. The notations are the same as in FIG. 1}
\end{figure}
\begin{figure}[h!]
  \centering
\includegraphics[scale=0.4]{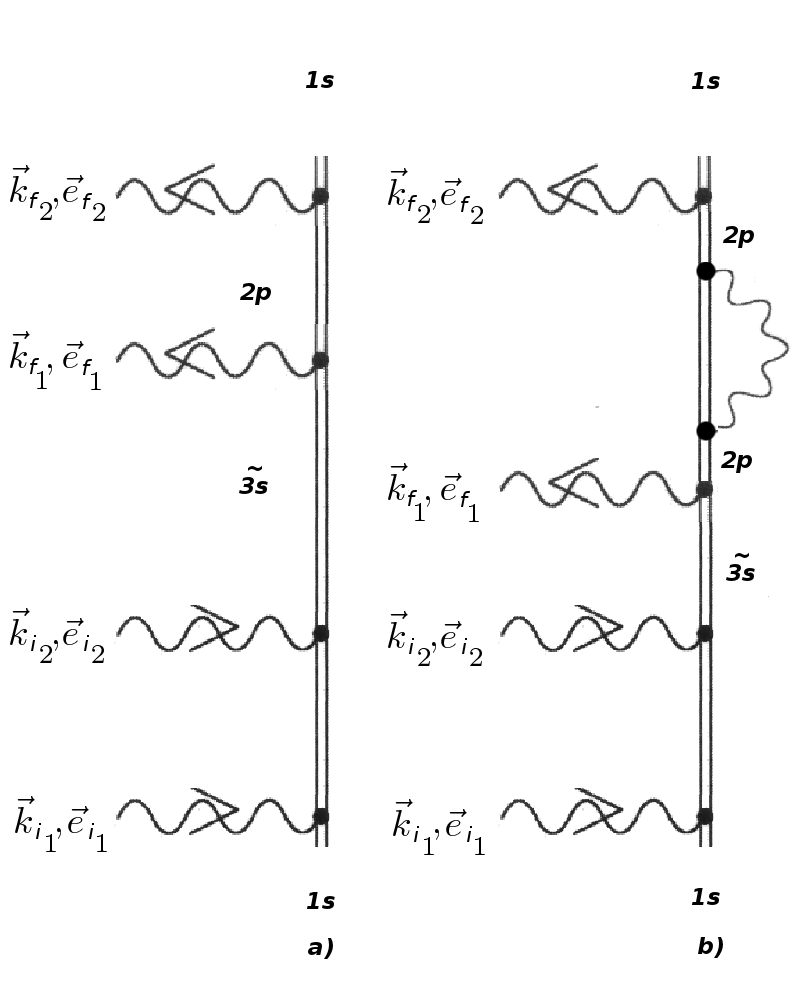}
  \caption{Feynman graph describing the two-photon resonance scattering on the ground state of hydrogen atom, with the excitation of $ 3s $ state (resonance condition $ \omega_{i_1}+\omega_{i_2}=E_{3s}-E_{1s} $ and the decay cascade resonances $ \omega^{res.1}=E_{3s}-E_{2p} $, $ \omega^{res.2}=E_{2p}-E_{1s} $. In FIG. 3$ a $ the basic process of the resonant scattering with the excitation of the $ 3s $ level and decay $ 3s-2p-1s $ is depicted. In FIG. 3$ b $ the electron self-energy insertion in the upper electron propagator is made. Notation $ \tilde{3s} $ means that the Low procedure is already performed for this electron line. The other notations are the same as FIGS. 1, 2}
\end{figure}
\begin{figure}[h!]
  \centering
\includegraphics[scale=0.4]{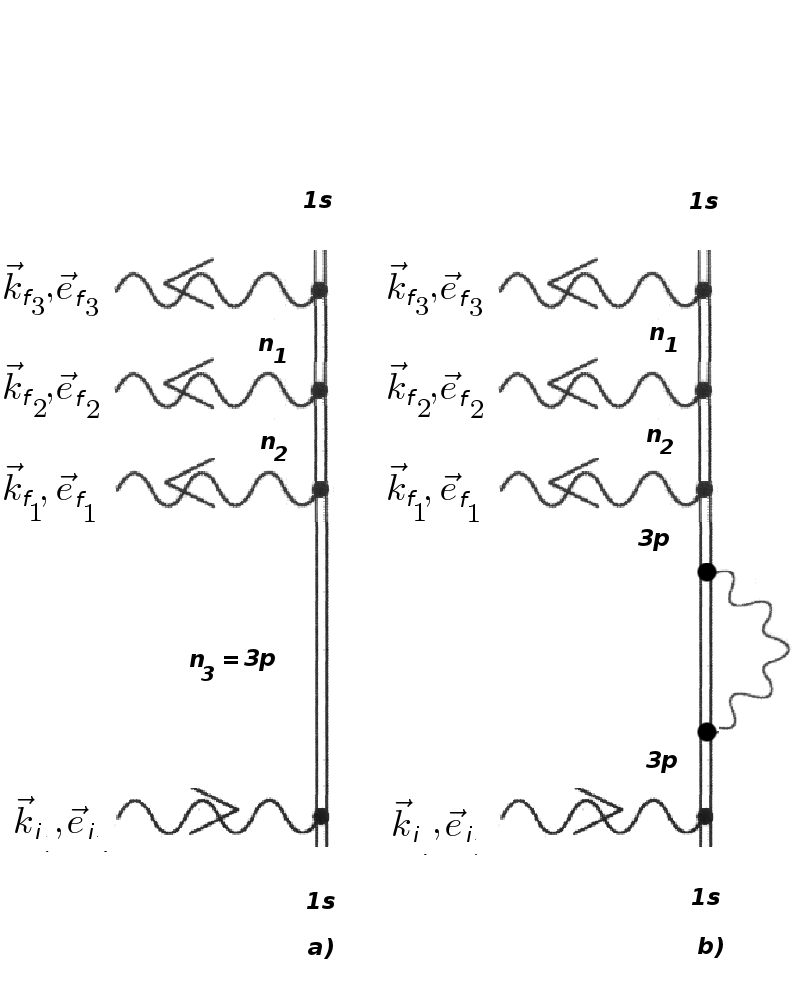}
  \caption{Feynman graph describing the resonant process with one-photon absorption and three-photon emission for the ground state of the hydrogen atom. In FIG. 4$ a $ the basic process is shown and in FIG. 4$ b $ the electron self-energy insertion in the central electron propagator is made. The notations are the same as FIG. 1}
\end{figure}
\end{document}